# Electromagnetic radiation and the self field of a spherical dipole oscillator

Masud Mansuripur† and Per K. Jakobsen‡

†College of Optical Sciences, The University of Arizona, Tucson

‡Department of Mathematics and Statistics, UIT The Arctic University of Norway, Tromsø, Norway

**Abstract**. For an oscillating electric dipole in the shape of a small, solid, uniformly-polarized, spherical particle, we compute the self-field as well as the radiated electromagnetic field in the surrounding free space. The assumed geometry enables us to obtain the exact solution of Maxwell's equations as a function of the dipole moment, the sphere radius, and the oscillation frequency. The self field, which is responsible for the radiation resistance, does *not* introduce acausal or otherwise anomalous behavior into the dynamics of the bound electrical charges that comprise the dipole. Departure from causality, a well-known feature of the dynamical response of a charged particle to an externally applied force, is shown to arise when the charge is examined in isolation, namely in the absence of the restraining force of an equal but opposite charge that is inevitably present in a dipole radiator. Even in this case, the acausal behavior of the (free) charged particle appears to be rooted in the approximations used to arrive at an estimate of the self-force. When the exact expression of the self-force is used, our numerical analysis indicates that the impulse-response of the particle should remain causal.

**1. Introduction**. Classical electrodynamics, an elegant theory based on Maxwell's equations and the Lorentz force law, is a wide-ranging and self-consistent physical theory that is also consistent with special relativity and with the principles of conservation of energy as well as linear and angular momenta.[1-6] The theory, however, runs into trouble when attempting to explain the action of small, point-like charged particles upon themselves.[7-11] An accelerated charged particle radiates an electromagnetic (EM) field, the action of which on the particle itself could, according to the classical theory, elicit an acausal response from the particle. The particle may thus be required to behave as if it were reacting or responding to an external excitation *before* the onset of that excitation. In the early years of the twentieth century, Max Abraham[7] and Hendrik Lorentz[8] studied the self-action of an accelerated, electrically-charged particle in the shape of a small sphere, and pointed out the possibility of its acausal behavior. Some thirty years later, Paul Dirac[11] analyzed the relativistic version of the same problem, found a clever way to eliminate the troublesome infinities that had previously hampered the investigations of point particles, and derived an exact solution for the self-force of an accelerated point charge. Unfortunately (for the classical theory), Dirac's exact solution exhibits the same anomaly of runaway solutions and causality violation that the approximate Abraham-Lorentz theory had previously encountered.

With the advent of quantum mechanics and the recognition that Heisenberg's uncertainty principle forbids the simultaneous specification of the position and momentum of small atomic and sub-atomic particles, it was hoped that the aforementioned foundational problems of the classical theory could be resolved, and that quantum electrodynamics would provide a satisfactory answer to such vexing problems as the violation of causality by an accelerated point charge.[1,2] Although substantial progress has been made since the formulation of quantum electrodynamics and a number of classical puzzles have been resolved, the problems associated with the self-force of an accelerated point-charge continue to attract the attention of theoretical physicists to this day.[12-30] Much has been written about these problems in textbooks,[1-6] review papers,[13-19] monographs,[20,21] and research articles[22-29] (to cite only a few), thus making it unnecessary here to discuss the history and the current state of affairs in any great detail. The interested reader can find a good overview of the subject in J. D. Jackson's *Classical Electrodynamics* (Ref. [2], Chapter 16). The monograph by A. D. Yaghjian[21] is an invaluable resource for in-depth understanding of the electrodynamics of charged spheres. Steane[19] describes the pathological behavior exhibited by certain equations of motion in the presence of self-force, and examines a class of formulations that do not show such pathologies. For an overview of attempts to mitigate or eliminate the runaway solutions and/or the predicted acausal behavior of charged particles in the presence of radiation reaction (either according to the Abraham-Lorentz theory, or due to the self-force in Dirac's fully relativistic formulation), the reader is referred to Rohrlich.[14,15,20]

Our rather narrow goal in the present paper is to examine a special case of the Abraham-Lorentz problem for which an exact analytical expression for the self-force can be obtained. When the approximate form of this self-force is used to derive the response of the particle to an externally applied impulsive excitation, we find that the emerging acausal behavior is the same as that predicted by the Abraham-Lorentz theory. However, a numerical analysis of the same problem that takes into account our *exact* expression of the self-force, indicates that the particle's impulse-response is causal — even when the radius of the particle assumes extremely small values. The conclusion, in agreement with the current understanding of the pathologies associated with small charged particles,[14] is that the acausal behavior predicted by the Abraham-Lorentz theory for a charged particle of finite size may not herald a failure of the Maxwell-Lorentz electrodynamics, but rather be an artifact of the approximations made to arrive at an estimate of the self-force. Our results, however, do not contradict the existence of runaway solutions or the acausal behavior of point charges as predicted by Dirac's theory,[11] since Dirac's self-force pertains to a particle of zero-size (i.e., a true point-particle), whereas our analysis applies to particles of finite (albeit very small) dimensions.[30]

The organization of the paper is as follows. After a brief synopsis in Sec.2, we solve Maxwell's equations in Sec.3 for a uniformly-polarized solid sphere of radius $R$, whose polarization $\boldsymbol{P}(t)$ oscillates at a constant frequency $\omega$. Here we derive the electric and magnetic fields both inside and outside the sphere and, among other things, find an exact formula for the electric field that is responsible for the radiation resistance. In Sec.4, the damping effect of this self-field is incorporated into the Lorentz oscillator model[1,2] of the spherical dipole, where we show that well-known classical results such as broadening and shifting of the resonant line-shape[2] and the Thomson scattering cross-section of the dipole[2] emerge when an approximate form of the radiation reaction function is used in the model.

We switch gears in Sec.5 and use the exact frequency-dependent transfer function of the Lorentz oscillator obtained in Sec.4 to examine the response of the dipole to an impulsive excitation. This requires a numerical analysis of the poles of the transfer function in the complex $\omega$-plane. The exact radiation reaction function will be seen to endow the transfer function with an infinite number of poles, none of which appears to reside in the upper half of the $\omega$-plane. We track the evolution of these poles by following their $\omega$-plane trajectories as functions of the dipole radius $R$, and find them to remain in the lower half-plane — even when $R$ becomes exceedingly small (i.e., below the femtometer scale of the classical electron radius[†]). The conclusion is that the impulse-response of the dipole is going to be causal, whether the dipole is large (i.e., $R \sim 1$ nm), or has the typical dimensions of a hydrogen atom ($R \sim 1$Å), or becomes as small as a nuclear particle ($R \sim 1$ fm), or when its radius assumes even smaller values.

Finally, in Sec.6, we relax the constraints of the Lorentz oscillator model on the ball of negative charge by eliminating the restraining force of the dipole's positive charge as well as that of the phenomenological spring that is built into the Lorentz oscillator model. The free ball of charge now responds to an impulsive excitation in a way that parallels the behavior predicted by the Abraham-Lorentz theory when we use an approximate form of the radiation reaction force. (Specifically, the free charged particle responds in an acausal manner to the impulsive force.) However, when the model incorporates the *exact* radiation reaction force, we find that the poles of the transfer function do not leave the lower half of the $\omega$-plane, indicating that the impulse-response is causal no matter how small a value is assumed for the radius $R$ of the charged particle. Once again, we find that the predicted acausal behavior reported in the older literature is likely a consequence of the approximations used to estimate the self-force, and that the exact solution of Maxwell's equations yields a causal impulse response — down to extremely small radii of the charged particle that may even go below the classical electron radius.

---

[†]Let a spherical shell of radius $r_c$ and charge $q$, where the charge is uniformly distributed over the sphere's surface, be a model for a stationary electron. Upon integration over the entire space, the $E$-field energy-density $\tfrac{1}{2}\varepsilon_0 E^2(r) = q^2/(32\pi^2 \varepsilon_0 r^4)$ outside the shell yields the total EM energy of the electron as $\mathcal{E} = q^2/(8\pi\varepsilon_0 r_c)$. Equating $\mathcal{E}$ to the mass-energy $m_0 c^2$, one obtains the *classical diameter* of the electron as $2r_c = \mu_0 q^2/(4\pi m_0) \cong 2.818$ fm.

**2. Synopsis**. This paper presents the special case of a spherical dipole for which an exact expression for the self-field (i.e., the electric field responsible for radiation resistance) can be found. This is done by a straightforward (albeit tedious) solution of the equations of classical electrodynamics presented in Appendices A and B. To our knowledge, the particular expression for the self-field of an oscillating dipole appearing in Sec.3, Eq.(13), has not been reported in the extant literature. In contrast to the well-known Abraham-Lorentz-Dirac (ALD) equation that is the usual point of departure in traditional discussions of charged-particle dynamics,[14,15,19,20] our self-field does *not* yield an equation of motion that is readily recognizable as a finite-order differential equation. Incorporating this exact expression of the self-field into the Lorentz oscillator model enables us to view the classical Abraham-Lorentz problem from a somewhat different perspective, one in which the response of a solid, uniformly-charged, non-deformable spherical ball to an impulsive excitation can be examined with and without the small-radius approximations.

Rohrlich[14] has discussed the case involving a similarly exact solution of the Maxwell-Lorentz equations for a spherical shell of uniform surface charge, stating that "the case of a volume charge is considerably more complicated and adds nothing to the understanding of the problem." Be it as it may, we believe it is worthwhile to bring to the community's attention the existence of an exact expression for the self-force of a solid, uniformly-charged sphere under the conditions reported in the following sections — if for no other reason than to ensure that the neglect of "nonlinear terms" that is inherent to the otherwise "exact" solution discussed by Rohrlich does not adversely affect his conclusions.

In another departure from the conventional approach, we do *not* renormalize the mass of the charged particle, keeping the full expression of the self-force throughout our analysis. Traditionally, renormalization is done by subtracting a term proportional to $1/R$ from the observed inertial mass $m_0$ of the particle, yielding what is known as the bare mass.[14] In our case, this would require subtracting $\mu_0 q^2/5\pi R$ from $m_0$; see the coefficient of $\omega^2$ in the denominator of the approximate transfer function given in Eq.(21). Our main reason for setting mass renormalization aside is that we are not convinced that the coefficient of $\omega^2$ should carry the entire burden of accounting for the contribution of electrodynamic mass to the particle inertia and, consequently, would like to postpone any discussion of mass renormalization until such time as we have a better grasp of the role played by $m_0$ in the charged-particle dynamics.

In the meantime, the absence of mass renormalization from our analysis should have very little effect as long as the radius $R$ is somewhat greater than the classical radius $r_c$ of the particle. As $R$ approaches $r_c$ from above and then goes below this critical radius, mass renormalization (in one form or another) should no longer be ignored, but at this point we are already deep inside the non-classical territory. According to Rohrlich,[14] "*the **classical** equations of motion have their validity limits where quantum mechanics becomes important: they can no longer be trusted at distances of the order of (or below) the Compton wavelength.*" (The Compton wavelength of the electron is $\lambda = h/m_0 c \cong 2.426$ pm.) Thus, in the regime where $R \lesssim r_c$, our analysis of the charged-particle's impulse-response should only be taken at face value since, deep inside the non-classical regime and absent a reliable understanding of the role played by the inertial mass $m_0$, it is undeniable that any purely mathematical result is devoid of physical content.[‡]

In a nutshell, the existing literature pertaining to the Abraham-Lorentz problem and relying solely on the ALD equation contends that a small ball of charge exhibits runaway solutions and that, in response to external excitations, it could behave in acausal fashion. Such pathologies in the predicted behavior of the particle disappear when more accurate estimates of the self-force are used in its equation of motion, with the caveat of taking the bare mass to be non-negative.[14] (For an electron, this constraint on the bare mass translates into $R \gtrsim r_c$.) The results reported in the present paper similarly indicate that, (i) when small-radius approximations are invoked, the predicted acausal behavior parallels those reported in the literature in accordance with the ALD equation, and (ii) our exact solution, examined numerically and down to very small particle radii shows no such acausal behavior. Thus, our exact results for uniformly-charged solid spheres agree with those in the literature for uniformly-charged spherical shells.[14] One potentially important difference is that our exact expression of the self-force does not suffer from the neglect of the aforementioned nonlinear terms.[14] Another difference is that our predicted causal behavior (for the bound charge within the dipole of Sec.5 as well as the free particle of Sec.6) extends to small solid spheres well below the classical electron diameter of ~2.818 fm — although this finding may not survive if and when we find a proper mass-renormalization scheme and, in any case, such classical speculations deep inside a non-classical regime are bereft of physical value.

**3. The electromagnetic field inside and outside an oscillating spherical dipole.** Consider an electric dipole in the form of a small sphere of radius $R$ and uniform polarization $P_0 \hat{z} \cos(\omega_s t)$, sitting at the origin of coordinates and oscillating at the source frequency $\omega_s$. Working in the spherical coordinate system $\mathbf{r} = (r, \theta, \varphi)$, and defining the function sphere$(r)$ to be zero when $r > 1$ and 1.0 when $r \leq 1$, the polarization distribution may be written as

$$\mathbf{P}(\mathbf{r},t) = P_0 \hat{z} \cos(\omega_s t) \, \text{sphere}(r/R).^{\S} \tag{1}$$

The particle's dipole moment is readily seen to be $\mathbf{p}_0(t) = (4\pi R^3/3) P_0 \cos(\omega_s t) \hat{z}$. The first step in evaluating the EM field that emanates from the oscillating dipole is to compute the four-dimensional Fourier transform $\mathbf{P}(\mathbf{k}, \omega)$ of $\mathbf{P}(\mathbf{r}, t)$. Appendix A shows that this Fourier transformation yields

$$\mathbf{P}(\mathbf{k}, \omega) = 3\pi p_0 \hat{z} [\delta(\omega + \omega_s) + \delta(\omega - \omega_s)] \, [\sin(kR) - kR\cos(kR)]/(kR)^3. \tag{2}$$

In classical electrodynamics, the bound electric charge and current densities are given by[2-6]

$$\rho^{(e)}_{\text{bound}}(\mathbf{r},t) = -\nabla \cdot \mathbf{P}(\mathbf{r},t) \qquad \rightarrow \qquad \rho^{(e)}_{\text{bound}}(\mathbf{k},\omega) = -\mathrm{i}\mathbf{k} \cdot \mathbf{P}(\mathbf{k},\omega). \tag{3}$$

$$\mathbf{J}^{(e)}_{\text{bound}}(\mathbf{r},t) = \partial \mathbf{P}(\mathbf{r},t)/\partial t \qquad \rightarrow \qquad \mathbf{J}^{(e)}_{\text{bound}}(\mathbf{k},\omega) = -\mathrm{i}\omega \mathbf{P}(\mathbf{k},\omega). \tag{4}$$

---

[‡] As a minor solace, one might argue that, if the goal is to show that the classical physics of a charged particle remains causal when the particle radius shrinks to extremely small (but nonzero) values, then it is perhaps advisable to stay away from the conventional — and arguably non-classical — stratagem of mass renormalization.

[§] In the literature, sphere$(r/R)$ is sometimes written as the Heaviside step function $\Theta(R - r)$.

In the absence of free charges, free currents, and magnetization, the bound charge and current densities of Eqs.(3) and (4) will be the total charge and current densities, which are directly related to the scalar potential $\psi(r,t)$ and vector potential $A(r,t)$, as follows:[2-6]

$$A(r,t) = \frac{\mu_0}{(2\pi)^4} \int_{-\infty}^{\infty} \frac{J(k,\omega)}{k^2-(\omega/c)^2} \exp[i(k \cdot r - \omega t)] \, dk d\omega. \qquad (5)$$

$$\psi(r,t) = \frac{1}{(2\pi)^4 \varepsilon_0} \int_{-\infty}^{\infty} \frac{\rho(k,\omega)}{k^2-(\omega/c)^2} \exp[i(k \cdot r - \omega t)] dk d\omega. \qquad (6)$$

A detailed step-by-step calculation of these potentials is relegated to Appendix B, where the final results for the regions inside and outside the spherical dipole are given in Eqs.(B17)-(B20). Subsequently, the scalar and vector potentials are used to determine the electric and magnetic fields inside and outside the dipole using the standard formulas[1-6]

$$E(r,t) = -\nabla \psi(r,t) - \partial A(r,t)/\partial t, \qquad (7)$$

$$B(r,t) = \nabla \times A(r,t). \qquad (8)$$

In the region outside the sphere of radius $R$, the $E$ and $B$ fields are found to be

$$E_{\text{out}}(r,t) = -\left(\frac{3p_0}{4\pi\varepsilon_0 r^3}\right)\left[\frac{\sin(R\omega_s/c)-(R\omega_s/c)\cos(R\omega_s/c)}{(R\omega_s/c)^3}\right]\{(r\omega_s/c)^2 \cos[\omega_s(t-r/c)] \sin\theta\, \widehat{\theta}$$
$$+ \{(r\omega_s/c)\sin[\omega_s(t-r/c)] - \cos[\omega_s(t-r/c)]\}(2\cos\theta\, \hat{r} + \sin\theta\, \widehat{\theta})\}. \qquad (9)$$

$$B_{\text{out}}(r,t) = -\left(\frac{3\mu_0 \omega_s p_0 \sin\theta\, \widehat{\varphi}}{4\pi r^2}\right)\left[\frac{\sin(R\omega_s/c)-(R\omega_s/c)\cos(R\omega_s/c)}{(R\omega_s/c)^3}\right]$$
$$\times \{(r\omega_s/c)\cos[\omega_s(t-r/c)] + \sin[\omega_s(t-r/c)]\}. \qquad (10)$$

The EM fields inside the particle are given by

$$E_{\text{in}}(r,t) = -\left(\frac{3p_0 \hat{z}}{4\pi\varepsilon_0 R^3}\right)\cos(\omega_s t)$$
$$+ \left(\frac{3p_0}{4\pi\varepsilon_0 R^3}\right)\left[\frac{\sin(r\omega_s/c)-(r\omega_s/c)\cos(r\omega_s/c)}{(r\omega_s/c)^3}(2\cos\theta\, \hat{r}+\sin\theta\, \widehat{\theta}) - \frac{\sin(r\omega_s/c)}{(r\omega_s/c)}\sin\theta\, \widehat{\theta}\right]$$
$$\times \{\cos[\omega_s(t-R/c)] - (R\omega_s/c)\sin[\omega_s(t-R/c)]\}. \qquad (11)$$

$$B_{\text{in}}(r,t) = -\left(\frac{3\mu_0 \omega_s p_0 \sin\theta\, \widehat{\varphi}}{4\pi r^2}\right)\left\{\frac{[\cos(R\omega_s/c)+(R\omega_s/c)\sin(R\omega_s/c)]\sin(\omega_s t)}{(R\omega_s/c)^3}\right.$$
$$\left. - \frac{[\sin(R\omega_s/c)-(R\omega_s/c)\cos(R\omega_s/c)]\cos(\omega_s t)}{(R\omega_s/c)^3}\right\}[\sin(r\omega_s/c) - (r\omega_s/c)\cos(r\omega_s/c)]. \quad (12)$$

The spatially averaged $E$-field inside the dipole may now be computed, as follows:

$$\langle E_{\text{in}}(r,t)\rangle = \left(\frac{3}{4\pi R^3}\right)\int_{r=0}^{R}\int_{\theta=0}^{\pi} E_{\text{in}}(r,t) 2\pi r^2 \sin\theta\, dr d\theta$$
$$= \left(\frac{3p_0 \hat{z}}{4\pi\varepsilon_0 R^3}\right)\left\{\frac{2[\sin(R\omega_s/c)-(R\omega_s/c)\cos(R\omega_s/c)] \times \{\cos[\omega_s(t-R/c)]-(R\omega_s/c)\sin[\omega_s(t-R/c)]\}}{(R\omega_s/c)^3} - \cos(\omega_s t)\right\}$$
$$= \left(\frac{p_0 \hat{z}}{\varepsilon_0}\right)\text{Re}\left\{\left[\frac{2[\sin(R\omega_s/c)-(R\omega_s/c)\cos(R\omega_s/c)] \times [1-i(R\omega_s/c)]e^{iR\omega_s/c}}{(R\omega_s/c)^3} - 1\right]\exp(-i\omega_s t)\right\}. \qquad (13)$$

In deriving Eq.(13), we used the symmetry of Eq.(11), which dictates that the components of $\boldsymbol{E}_{\text{in}}(\boldsymbol{r},t)$ parallel to the $xy$-plane average out to zero. When integrated over $\theta$, the projection along $\hat{\boldsymbol{z}}$ of $(2\cos\theta\,\hat{\boldsymbol{r}} + \sin\theta\,\widehat{\boldsymbol{\theta}})$ yields $\int_0^\pi (2\cos^2\theta - \sin^2\theta)\sin\theta\,d\theta = 0$, whereas that of $\sin\theta\,\widehat{\boldsymbol{\theta}}$ yields $-\int_0^\pi \sin^3\theta\,d\theta = -4/3$. We also used the fact that $\int_0^a x\sin x\,dx = \sin a - a\cos a$. Note that the Lorentz force exerted on the dipole's moving charge by the magnetic field $\boldsymbol{B}_{\text{in}}(\boldsymbol{r},t)$ is everywhere parallel to the $xy$-plane and radially symmetric with respect to the $z$-axis. Consequently, there cannot be any radiation reaction forces due to this (internal) magnetic field.

In the last line of Eq.(13), the term inside the square brackets can be expanded in powers of $\zeta = R\omega_s/c$ to yield

$$\frac{2(\sin\zeta - \zeta\cos\zeta)(1-i\zeta)e^{i\zeta}}{\zeta^3} - 1 = -\tfrac{1}{3} + \tfrac{4}{15}\zeta^2 + \tfrac{2i}{9}\zeta^3 - \tfrac{4}{35}\zeta^4 + \cdots. \tag{14}$$

For typical atomic dipoles, the condition $R \ll \lambda_s = 2\pi c/\omega_s$ implies that $\zeta = R\omega_s/c \ll 1$. Therefore, the dominant contribution to the average internal $E$-field of Eq.(13) comes from the static term $-P_0\hat{\boldsymbol{z}}/3\varepsilon_0$. The third term on the right-hand side of Eq.(14) is primarily responsible for radiation damping (or radiation resistance).

Outside the dipole, the time-averaged Poynting vector $\langle \boldsymbol{S}(\boldsymbol{r},t)\rangle$ is computed as follows:

$$\langle \boldsymbol{S}(\boldsymbol{r},t)\rangle = \langle \boldsymbol{E}_{\text{out}} \times \boldsymbol{H}_{\text{out}}\rangle = \left(\frac{3p_0}{4\pi R^3}\right)^2 \left[\frac{\omega_s \sin^2\theta}{2\varepsilon_0(\omega_s/c)^3 r^2}\right][\sin(R\omega_s/c) - (R\omega_s/c)\cos(R\omega_s/c)]^2 \hat{\boldsymbol{r}}. \tag{15}$$

The total radiated power is obtained by integrating the above $\langle \boldsymbol{S}(\boldsymbol{r},t)\rangle$ over a spherical surface of radius $r$. We find

$$\text{emitted power} = \left(\frac{3p_0}{4\pi R^3}\right)^2 \frac{4\pi\omega_s}{3\varepsilon_0(\omega_s/c)^3}[\sin(R\omega_s/c) - (R\omega_s/c)\cos(R\omega_s/c)]^2 \cong \frac{\mu_0 p_0^2 \omega_s^4}{12\pi c}. \tag{16}$$

The approximate form in Eq.(16) is the result of small-angle approximations, $\sin x \cong x - x^3/3!$ and $\cos x \cong 1 - x^2/2!$, which yield $\sin(R\omega_s/c) - (R\omega_s/c)\cos(R\omega_s/c) \cong \tfrac{1}{3}(R\omega_s/c)^3$. The total radiated power is thus seen to be proportional to the square of the dipole moment $p_0$ and, in its approximate form, proportional to the fourth power of the oscillation frequency $\omega_s$.

**4. The Lorentz oscillator model.** With reference to Fig.1, we imagine the dipole as consisting of two overlapping balls of radius $R$ and uniformly distributed positive and negative charges $\pm q$, with the positive ball being massive and, therefore, immobile, while the negative ball of mass $m_0$ is free to move (ever so slightly) up and down along the $z$-axis. A short spring, having spring constant $\alpha$, connects the centers of the two balls and exerts a restoring force on the negatively-charged ball along the $z$ direction. A dynamic friction coefficient $\beta$ accounts for the internal losses due to a force that is assumed to be proportional to the instantaneous velocity of the oscillating ball.[1,2] The only other forces acting on the negatively-charged ball are due to an externally applied uniform electric field $\boldsymbol{E}_{\text{ext}}(t) = E_0 \cos(\omega t)\hat{\boldsymbol{z}}$, and the (spatially-averaged) internal dipolar $E$-field $\langle \boldsymbol{E}_{\text{in}}(\boldsymbol{r},t)\rangle$ given by Eq.(13). The spatial averaging of the internal $E$-field is justified by our earlier assumption that both spheres are rigidly and uniformly charged.

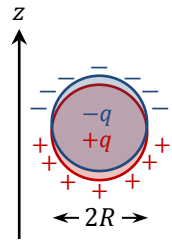

**Fig.1**. A pair of solid spherical balls of identical radius $R$ overlap each other in 3-dimensional space. The spheres are uniformly filled with electric charge, one containing a total charge of $+q$, the other a total charge of $-q$. The positively-charged ball is massive and immobile, whereas the negative ball, having a finite inertial mass $m_0$, can oscillate along the $z$-axis, its minute displacement from the equilibrium position being denoted by $z(t)$. The electric dipole moment thus produced is $\boldsymbol{p}_0(t) = -qz(t)\hat{\boldsymbol{z}}$. Sinusoidal motion of the negative ball gives rise to the time-dependent dipole moment $p_0\hat{\boldsymbol{z}}\cos(\omega t)$, which radiates an EM field of frequency $\omega$ into the surrounding free space. Also created is an oscillatory EM field inside the negatively-charged sphere that is responsible for the radiation resistance.

The critical underlying assumptions in our model of the dipole are that the two inter-penetrating balls of charge depicted in Fig.1 are solid (i.e., each occupies the entire volume of its corresponding sphere), are non-deformable and that, within their respective volumes, they have a uniform charge-density distribution. These are essential assumptions that allow one to find the self-field of radiation resistance as the *spatial average* of the internal $E$-field given by Eq.(13). It is true, of course, that no material object can exhibit infinite rigidity in view of the constraints of special relativity; nevertheless, our assumption of particle non-deformability is in keeping with the underlying hypotheses of the Abraham-Lorentz theory, at least to the extent that it pertains to the structure of sub-atomic particles.

Denoting by $z(t)$ the displacement of the negatively-charged ball along the $z$-axis, we may now write the Newtonian law of motion[**] for this ball as follows:

$$m_0 \frac{d^2 z(t)}{dt^2} = -qE_0 e^{-i\omega t} - q\langle E_{\text{in}}(\mathbf{r}, t)\rangle - \alpha z(t) - \beta \frac{dz(t)}{dt}. \tag{17}$$

The induced dipole is thus given by $\mathbf{p}(t) = -qz(t)\hat{\mathbf{z}} = p_0 \hat{\mathbf{z}} e^{-i\omega t}$. Defining the normalized parameters $\omega_0 = \sqrt{\alpha/m_0}$ (resonance frequency) and $\gamma = \beta/m_0$ (damping coefficient), while denoting the charged ball's volume by $v = 4\pi R^3/3$, and the bracketed term in the last line of Eq.(13) by $\Gamma(\omega)e^{-i\omega t}$, the streamlined version of Eq.(17) becomes

$$\frac{d^2}{dt^2}(p_0 e^{-i\omega t}) = \left(\frac{q^2}{m_0}\right) E_0 e^{-i\omega t} + \left(\frac{q^2}{m_0}\right)\left(\frac{p_0}{\varepsilon_0 v}\right) \Gamma(\omega) e^{-i\omega t} - \omega_0^2 p_0 e^{-i\omega t} - \gamma \frac{d}{dt}(p_0 e^{-i\omega t}). \tag{18}$$

The so-called "radiation reaction" function $\Gamma(\omega)$ appearing in the preceding equation is given by

$$\Gamma(\omega) = \frac{2[\sin(R\omega/c) - (R\omega/c)\cos(R\omega/c)] \times [1 - i(R\omega/c)]e^{iR\omega/c}}{(R\omega/c)^3} - 1. \tag{19}$$

Recall that the first few terms in the Taylor series expansion of $\Gamma(\omega)$ in terms of $\zeta = R\omega/c$ are listed in Eq.(14). Introducing the plasma frequency $\omega_p = q/\sqrt{\varepsilon_0 m_0 v}$, and proceeding to simplify Eq.(18), we finally arrive at the transfer function relating the induced dipole amplitude $p_0$ to the applied $E$-field amplitude $E_0$, namely,

$$p_0 = \frac{q^2/m_0}{\omega_0^2 - \omega^2 - \omega_p^2 \Gamma(\omega) - i\gamma\omega} E_0. \tag{20}$$

Figure 2 shows computed profiles of (a) the absolute value, (b) the real part, and (c) the imaginary part of the transfer function $p_0/E_0$ of Eq.(20) versus the excitation frequency $\omega$ for a particle of radius $R = 1$ Å, charge $q = 1.6 \times 10^{-19}$ coulomb, and mass $m_0 = 9.11 \times 10^{-31}$ kg, corresponding to a spherical ball of charge roughly equivalent to a single electron in the 1$s$-state of the hydrogen atom. We have also set $\omega_0 = 3 \times 10^{15}$ rad/s to obtain a reasonable resonance frequency in the uv range of the optical spectrum, and $\gamma = 10^8$ rad/s for a relatively small contribution by non-radiative damping. (The constants of nature are $\varepsilon_0 \cong 8.854 \times 10^{-12}$ farad/m, $\mu_0 = 4\pi \times 10^{-7}$ henry/m, and $c = (\mu_0 \varepsilon_0)^{-\frac{1}{2}} \cong 3 \times 10^8$ m/s.)

In the case of sinusoidal excitation of the dipole, it suffices to use the first three terms in the Taylor series expansion of $\Gamma(\omega)$ to arrive at the following approximate transfer function:

---

[**]Since the displacement $z(t)$ of the negatively-charged sphere must be much less than the radius $R$ of the spherical particle, its velocity $V$ should be well below $R\omega$ and, therefore, $V/c \ll R\omega/c$. One can, therefore, push $R\omega/c$ to values as high as 10 before having to worry about the validity of Newton's (non-relativistic) equation of motion.

$$p_0 \cong \frac{q^2/m_0}{(\omega_0^2 + \tfrac{1}{3}\omega_p^2) - (1+\mu_0 q^2/5\pi m_0 R)\omega^2 - i(\mu_0 q^2/6\pi m_0 c)\omega^3 - i\gamma\omega} E_0. \tag{21}$$

However, to compute the dipole's impulse-response, it is necessary to use the exact form of $\Gamma(\omega)$, given in Eq.(19), in order to locate all the poles of the transfer function of Eq.(20) in the complex $\omega$-plane. We defer a discussion of the dipole's impulse-response to Sec.5, continuing for the time being with an examination of the approximate transfer function of Eq.(21).

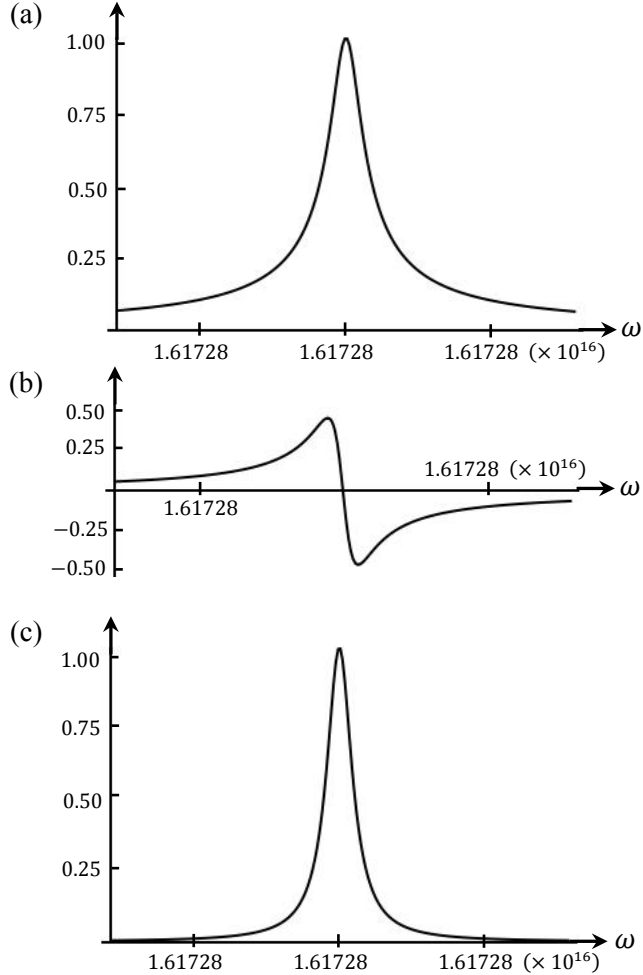

**Fig. 2**. Normalized profiles of (a) the absolute value, (b) the real part, and (c) the imaginary part of the transfer function $p_0/E_0$ of Eq.(20) versus the frequency $\omega$ for a spherical particle of radius $R = 1$ Å. The parameter values are chosen so that the negative ball of charge is roughly equivalent to a single electron in the 1s-state of the hydrogen atom. We have also set $\omega_0 = 3 \times 10^{15}$ rad/s to obtain a reasonable resonance frequency in the ultraviolet range of the optical spectrum, and $\gamma = 10^8$ rad/s for a relatively small contribution from non-radiative damping.

Recalling the Clausius-Mossotti correction[2,6] of the Lorentz oscillator model, one could argue that the term $\tfrac{1}{3}\omega_p^2$ added to $\omega_0^2$ in the denominator of Eq.(21) has already been taken into account when defining the spring constant $\alpha$. Leaving this argument aside, the effective resonance frequency is going to be $\widetilde{\omega}_0 = (\omega_0^2 + \tfrac{1}{3}\omega_p^2)^{½}$. Also, for typical atomic radii, the correction $\mu_0 q^2/(5\pi R)$ to the mass $m_0$ appearing in the coefficient of $\omega^2$ in Eq.(21) is quite small and, for all practical purposes, negligible. In the absence of internal loss mechanisms (e.g., absorption followed by non-radiative decay), one would set $\gamma$ to zero and rely solely on the radiation resistance term for damping. Under these circumstances, when the dipole is excited by a monochromatic plane wave of amplitude $E_0$ and frequency $\omega$, we will have

$$|p_0|^2 \cong \frac{(q^2/m_0)^2 E_0^2}{(\widetilde{\omega}_0^2 - \omega^2)^2 + (\mu_0 q^2/6\pi m_0 c)^2 \omega^6}. \qquad (22)$$

Defining the parameter $\tau = \mu_0 q^2/(6\pi m_0 c)$, it is not difficult to show that the resonance line-width is $\Delta\omega \sim \tau\widetilde{\omega}_0^2$, and that the resonance peak occurs at $\omega_{\text{peak}} \cong \widetilde{\omega}_0 - \frac{3}{4}\tau^2\widetilde{\omega}_0^3$. This estimate of the peak shift away from the resonance frequency $\widetilde{\omega}_0$, which is solely due to radiation reaction, is much smaller than typical atomic line-widths $\Delta\omega$. (One must resort to quantum mechanical treatments of atomic and molecular radiation to arrive at accurate and realistic estimates of both the resonance line-width and the corresponding peak shift.[2]) In the example depicted in Fig.2, using $v = 4\pi R^3/3$ for the particle volume, at $R = 1$Å we find $\omega_p = 2.753 \times 10^{16}$ rad/s and $\tau = 6.245 \times 10^{-24}$ s, yielding an effective resonance frequency $\widetilde{\omega}_0 = 1.617 \times 10^{16}$ rad/s and a line-width $\Delta\omega \sim 1.63 \times 10^9$ rad/s.

Using Eqs.(16) and (22), one can also compute the radiation emission rate per unit time. Considering that the rate of incident optical energy (per unit area per unit time) is $E_0^2/2Z_0$, where $Z_0 = (\mu_0/\varepsilon_0)^{\frac{1}{2}} \cong 377$ Ω is the impedance of free space, the dipole's scattering cross-section[2] is straightforwardly found to be

$$s \cong \frac{(\mu_0 q^2/m_0)^2 \omega^4}{6\pi[(\widetilde{\omega}_0^2 - \omega^2)^2 + (\mu_0 q^2/6\pi m_0 c)^2 \omega^6]}. \qquad (23)$$

In the above equation, the radiation resistance term is essentially negligible everywhere except in the vicinity of the resonance frequency $\omega \cong \widetilde{\omega}_0$, where it governs the peak value $s_{\text{max}}$ of the scattering cross-section and the width $\Delta\omega$ of the resonance peak, as follows:

$$s_{\text{max}} \cong 6\pi(c/\widetilde{\omega}_0)^2 = 3\lambda_0^2/2\pi, \qquad (24)$$

$$\Delta\omega \cong \mu_0 q^2 \widetilde{\omega}_0^2/(6\pi m_0 c). \qquad (25)$$

In accordance with Eq.(23), at frequencies well below resonance, we will have

$$s \cong (\mu_0 q^2/m_0)^2 (\omega/\widetilde{\omega}_0)^4/6\pi, \qquad (26)$$

and at frequencies well above resonance (but not extremely high), we will have

$$s \cong (\mu_0 q^2/m_0)^2/6\pi. \qquad (27)$$

All the results obtained thus far by approximating the exact radiation reaction function $\Gamma(\omega)$ of Eq.(19) are in complete accord with the classical results.[2] In the remaining part of the paper, we argue that any acausal behavior predicted by the classical theories is likely due to the approximate nature of those theories — the exception being Dirac's demonstration of runaway behavior based on his formula for the self-force of an accelerated point-charge, which is an exact relativistic solution of the Maxwell-Lorentz equations of classical electrodynamics.[11,30] Specifically, upon examining the impulse-response of the dipole in the next section using the exact radiation reaction function of Eq.(19), we do *not* find any indication of departure from causal behavior — even in the limit when the dipole radius $R$ becomes exceedingly small, going far below the classical radius $r_c$ of the electron. Similarly, in Sec.6, where we eliminate the restoring force of the positive charge as well as that of the (phenomenological) spring, and proceed to examine the impulse-response of a free-standing, negatively-charged sphere, we find acausal behavior only when the analysis relies on an approximate form of $\Gamma(\omega)$. In other words, the *exact* radiation reaction function of Eq.(19) acting on a small, electrically-charged sphere driven by an impulsive force, does *not* appear to give rise to acausal behavior.

**5. The impulse-response**. When excited by an external impulse, an initially dormant dipole exhibits a damped oscillatory response, with a time dependence that is precisely the inverse Fourier transform of the transfer function $p_0/E_0$ of Eq.(20). This is because the impulsive excitation consists of a uniform superposition of all sinusoidal frequencies $\omega$ from $-\infty$ to $\infty$. Consequently, the dipole's impulse-response is obtained by the inverse Fourier integral of the transfer function $p_0/E_0$ over the entire real-axis $\omega$. Now, according to Cauchy's theorem of complex analysis, when $t < 0$, the contour of integration can be closed with a large semi-circle in the upper half of the complex $\omega$-plane. The absence of poles in the upper-half plane would then imply that the inverse Fourier integral (i.e., the impulse-response) is zero when $t < 0$. This is the requirement for causal behavior that will be used throughout the following analysis. (We mention in passing that an impulsive excitation takes the dormant dipole instantaneously to an excited initial state, whence it decays to the ground state following the dynamical equation of motion of the dipole. Physically, this is tantamount to bringing an isolated atom to an excited state, then observing its decay to the ground state via spontaneous emission; see Appendix C for a discussion of the radiated EM energy when an initially dormant dipole is excited by an impulsive force.)

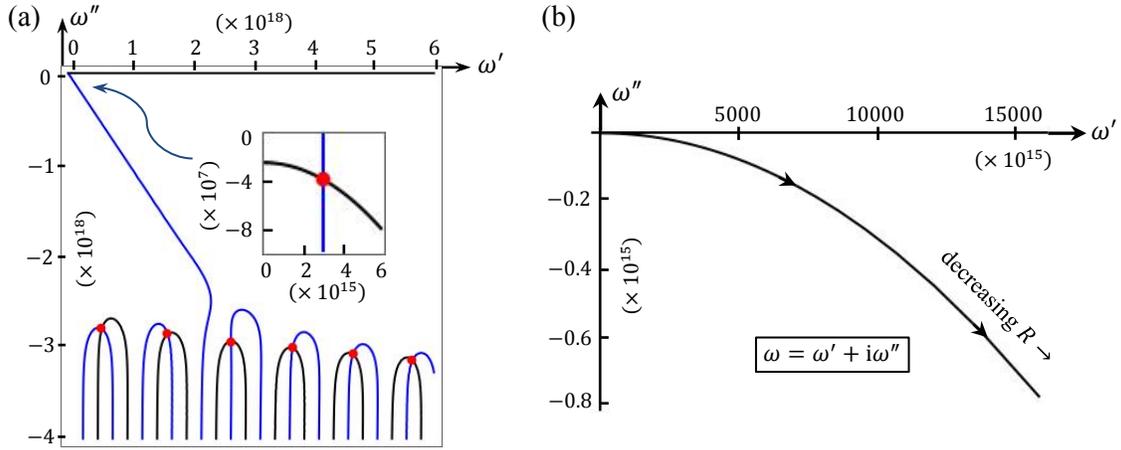

**Fig.3**. (a) Complex plane diagram showing, within the 4th quadrant of the $\omega$-plane, the zero contours of the real part (blue) and imaginary part (black) of the denominator of Eq.(20). A mirror image of these contours also resides in the 3rd quadrant. Here $R = 1.0$ nm, and the remaining parameters are the same as those used in Fig.2. The marked crossing points are the poles of the transfer function $p_0/E_0$. The inset is a magnified view of the region surrounding the dominant pole. (b) Trajectory of the dominant pole within the 4th quadrant of the $\omega$-plane. All the parameters are fixed except for the radius $R$ of the particle, which starts at 1.0 nm on the upper left-hand corner of the graph and goes down to 1.0 pm at its lower right-hand corner.

Causality, an important property of the impulse-response of the dipole, is governed by the location of the poles of its transfer function in the complex $\omega$-plane. Causal behavior is ensured if all the poles reside in the lower half of the $\omega$-plane. We undertook a detailed numerical study of the poles of the transfer function of Eq.(20) with the exact $\Gamma(\omega)$ of Eq.(19). Here, we describe the trajectories of these poles as the radius $R$ of the spherical dipole approaches zero, and confirm that, for all nonzero values of $R$, the poles remain in the lower half of the $\omega$-plane. Stated differently, we have found no indication that the dipole's impulse-response violates causality as the radius of the particle shrinks to extremely small values.

Figure 3(a) shows contours of zero real part (blue) and zero imaginary part (black) for the denominator $D(\omega)$ of the transfer function $p_0/E_0$ within the 4$^{th}$ quadrant of the complex $\omega$-plane. (A mirror image of these contours also resides in the 3$^{rd}$ quadrant.) The crossing points marked with red dots identify the poles of the transfer function. For the chosen set of parameters ($R = 1.0$ nm, $\omega_0 = 3 \times 10^{15}$ rad/s, $\gamma = 10^8$ rad/s, $\varepsilon_0$, $c$, $q$, and $m_0$ the same as those in Fig.2), the dominant pole, shown in the inset of Fig.3(a), is at $\omega \cong 3 \times 10^{15} - i\, 4 \times 10^7$ rad/s, whereas all the remaining poles are far below the real-axis.

Figure 3(b) shows the trajectory of the dominant pole as the radius $R$ of the particle declines from 1.0 nm at the upper left-hand corner of the graph down to 1.0 pm at the lower right-hand corner. At $R = 1.0$ pm, the dominant pole is located at $\omega = 1.5 \times 10^{19} - i\, 0.8 \times 10^{15}$ rad/s.

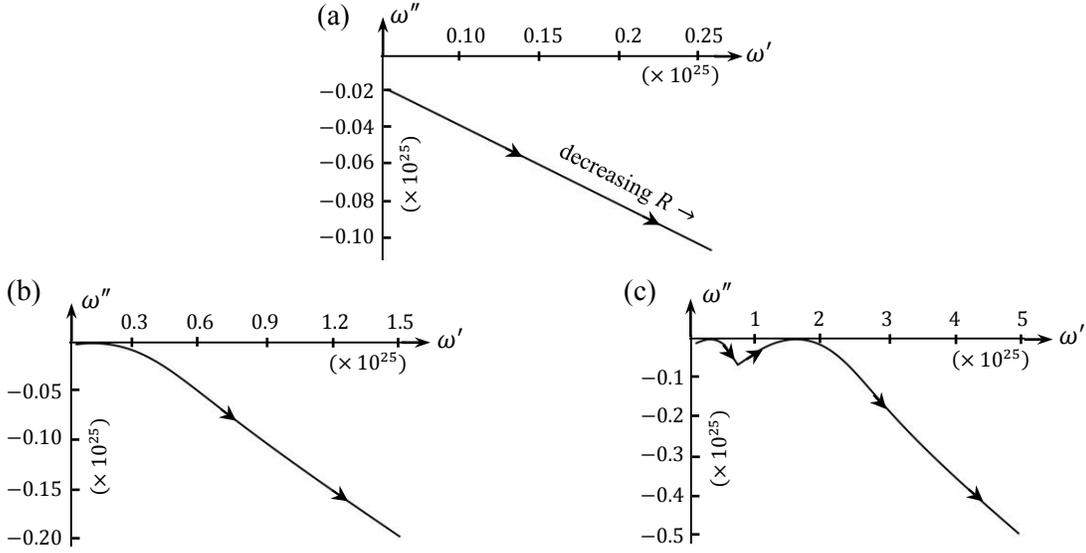

**Fig. 4**. $\omega$-plane trajectories of (a) the dominant pole, and (b, c) the 1$^{st}$ and 2$^{nd}$ distant poles depicted in Fig.3(a). In (a) and (b), $R$ is in the range of $[0.5, 0.1]$ fm; in (c) the range of $R$ is $[1.0, 0.05]$ fm.

Figure 4 shows an extended trajectory of the dominant pole as well as those of the first and second distant poles depicted in Fig.3(a) — i.e., poles identified with red dots, counting from left to right. The dominant pole is seen to continually move downward and to the right as $R$ drops from 0.5 fm to 0.1 fm. The first distant pole initially moves up toward the real axis (but never crosses it), then goes down and to the right. The second distant pole shows two initial humps (again, never crossing the real axis), before it drops down again.

We also set $\omega_0 = 0$ and $\gamma = 0$ to rule out the possibility that these phenomenological features of the Lorentz oscillator model might be responsible for the causal behavior of the impulse-response.[††] Under no circumstances did we observe any pole of the transfer function to move into the upper half of the $\omega$-plane. Application of Cauchy's argument principle with the aid of numerical integration (spanning the broad frequency range of $|\omega| \leq 3 \times 10^{33}$ rad/s) further affirmed these findings. We conclude that the dipole's response to an impulsive excitation must be causal, even when its radius $R$ assumes exceedingly small values, far below the classical electron radius.

---

[††]Ideally, when the radius $R$ becomes comparable to or smaller than the classical radius $r_c$ of the electron, one should also consider renormalizing the mass $m_0$ of the oscillating ball of charge. However, for the reasons mentioned in Sec.2, we believe that this subject is best left for future studies.

**6. Radiation reaction on a small spherical charge**. Returning to Eq.(17), if the driving force acting on the negatively-charged ball is $\boldsymbol{f}_{\text{ext}}(t) = -qE_0 \exp(-i\omega t)\hat{\boldsymbol{z}}$, and the small deviation from the ball's equilibrium position is written as $z(t) = z(\omega)\exp(-i\omega t)$, we will have

$$z(\omega) = \frac{f_{\text{ext}}(\omega)}{\alpha - m_0\omega^2 - (q^2/\varepsilon_0 v)\Gamma(\omega) - i\beta\omega}. \tag{28}$$

Now, the restoring force of the (stationary) positive ball on the (vibrating) negative ball is built into $\Gamma(\omega)$ as $\lim_{\omega \to 0} \Gamma(\omega) = -\tfrac{1}{3}$. Thus, by setting $\alpha = -q^2/(3\varepsilon_0 v)$, we eliminate the restoring force of the positive charge, leaving the negatively-charged ball as the only source of radiation reaction on itself. Also, setting $\beta = 0$ would eliminate the effects of non-radiative damping, but it is preferable to retain this friction coefficient for the time being. All in all, the oscillation amplitude $z(\omega)$ of a small solid sphere of radius $R$, charge $-q$, and mass $m_0$, in response to an externally-applied force $\boldsymbol{f}_{\text{ext}}(t) = f_0 \exp(-i\omega t)\hat{\boldsymbol{z}}$ — taking into account the radiation reaction force as well as a non-radiative damping force — will be

$$z(\omega) = -\frac{f_0(\omega)}{m_0\omega^2 + q^2[3\Gamma(\omega)+1]/(4\pi\varepsilon_0 R^3) + i\beta\omega}. \tag{29}$$

Invoking the Taylor series expansion (up to the 3$^{\text{rd}}$ order) of $3\Gamma(\omega) + 1 \cong \tfrac{4}{5}(R\omega/c)^2 + \tfrac{2}{3}i(R\omega/c)^3$ given in Eq.(14), one may approximate Eq.(29) as follows:

$$z(\omega) \cong -\frac{f_0(\omega)}{[m_0+(\mu_0 q^2/5\pi R)]\omega^2 + i(\mu_0 q^2/6\pi c)\omega^3 + i\beta\omega}. \tag{30}$$

Note in the above equation that the contribution of radiation reaction to the particle's mass is $\mu_0 q^2/(5\pi R)$, which, in the case of a single electron, approaches its inertial mass $m_0$ if $R$ happens to be in the vicinity of the classical electron radius (i.e., $\sim 10^{-15}$ meter). Let us denote by $m = m_0 + (\mu_0 q^2/5\pi R)$ the effective mass of the negatively charged ball, then normalize the remaining parameters by $m$, so that $\tau = \mu_0 q^2/(6\pi m c)$ and $\gamma = \beta/m$. The transfer function appearing in Eq.(30) has three poles at $\omega = 0$ and $\omega = \tfrac{1}{2}i(1 \pm \sqrt{1+4\gamma\tau})/\tau$. The response $z(t)$ of the negatively-charged sphere to the impulsive force $\boldsymbol{f}_{\text{ext}}(t) = F_0\delta(t)\hat{\boldsymbol{z}}$, computed with the aid of Cauchy's theorem and complex-plane integration, is readily found to be

$$z(t) = \left(\frac{F_0}{m}\right)\begin{cases} -\dfrac{1}{2\gamma} + \dfrac{2\tau \exp[(1+\sqrt{1+4\gamma\tau})t/2\tau]}{1+4\gamma\tau + \sqrt{1+4\gamma\tau}}; & t \leq 0, \\[1em] +\dfrac{1}{2\gamma} - \dfrac{2\tau \exp[(1-\sqrt{1+4\gamma\tau})t/2\tau]}{1+4\gamma\tau - \sqrt{1+4\gamma\tau}}; & t \geq 0. \end{cases} \tag{31}$$

Both $z(t)$ and its derivative $\dot{z}(t)$ are seen to be continuous at $t = 0$. Since Eq.(30) does not specify an initial condition (due to the presence of a pole at $\omega = 0$), the solution $z(t)$ can be augmented by an arbitrary constant $z_0$. Adding $z_0 = F_0/(2m\gamma)$ to Eq.(31) ensures that $z(-\infty) = 0$. We now let the damping coefficient $\gamma$ shrink to sufficiently small values to ensure that $\gamma\tau \ll 1$, in which case Eq.(31) becomes

$$z(t) \cong \left(\frac{F_0}{m}\right)\begin{cases} \tau\exp(t/\tau); & t \leq 0, \\[0.5em] \dfrac{1-\exp(-\gamma t)}{\gamma}; & t > 0. \end{cases} \tag{32}$$

The acausal behavior during $t < 0$, which is manifest in this nonzero response of the charged particle to an impulsive excitation at $t = 0$, has been deemed indicative of the failure of classical electrodynamics when applied to point particles.[19] Historically, this has been the lesson of the original Abraham-Lorentz theory pertaining to the effect of self-force on an accelerated point charge.[2] However, as pointed out in Sec. 2, more accurate calculations of the self-force for small charged particles have revealed that the acausal behavior disappears so long as the particle's bare mass remains positive — a condition that is equivalent to the particle radius being greater than the classical radius associated with its charge $q$ and inertial mass $m_0$.

When we used the exact $\Gamma(\omega)$ of Eq.(19) in Eq.(29), and conducted a numerical search for the poles of the transfer function in the $\omega$-plane, we found all the poles to reside in the lower half-plane. As was done in our numerical investigation of the impulse-response of a dipole in Sec.5, we allowed the radius $R$ of the solid sphere in the present case to shrink to exceedingly small values, and also examined situations in which $\gamma$ was set to zero. Under no circumstances did we find a pole in the upper half-plane. (These findings were further affirmed for values of $R$ as small as $10^{-25}$ fm by applying Cauchy's argument principle with the aid of numerical integration spanning the broad frequency range of $|\omega| \leq 3 \times 10^{48}$ rad/s.) The conclusion is that the acausal behavior exemplified by Eq.(32) is an artifact of the approximate nature of the radiation reaction force used in Eq.(30). In other words, when the transfer function incorporates the exact radiation reaction function of Eq.(19), the predicted behavior under all examined circumstances (even for particle radii far below their classical radius[‡‡]) remains causal.

**7. Concluding remarks**. We have examined a special case of EM radiation by an accelerated, uniformly-charged, non-deformable, solid sphere and, based on a numerical evaluation of the location of the poles of its transfer function, concluded that the response of the particle to an impulsive excitation must be causal. Our numerical investigation covered a broad range of particle radii from atomic-scale ($R \sim 1$ nm) down to sub-nuclear dimensions ($R \ll 1$ fm). We examined cases where the particle was confined within a dipole and, therefore, restrained by the attractive coulomb force of an equal but opposite charge, as well as cases where the restraining force of its opposite-charge partner was removed, thus releasing the particle from confinement. While these are special cases of accelerated charged particles, which do not merit blind generalization to other situations for which the Abraham-Lorentz type of analysis is pertinent — and definitely not cases to which Dirac's exact relativistic treatment applies — it is nonetheless important to recognize that, if not in all cases, then at least in some situations, acausal behavior is not inherent to the classical Maxwell-Lorentz theory, but rather is an artifact of the approximations used to estimate the self-force of an accelerated particle.

We did not automatically renormalize the inertial mass $m_0$ of our charged particle, not so much because its necessity is in doubt, but rather because we are not convinced that a simple reduction of $m_0$ by $\mu_0 q^2/5\pi R$ is the best way to account for the electrodynamic contributions to the particle inertia. We have, therefore, opted to postpone the issue of mass renormalization until such time as we have attained a better understanding for the role of $m_0$ in our equations of motion. In the meantime, for particle radii that are more or less greater than the critical radius $r_c$, our main conclusions are not significantly affected, given that electrodynamic contributions to the inertial mass in this regime are fairly small. For $R \lesssim r_c$, however, the conclusions could change, but here, as pointed out by Rohrlich,[14] we are already deep inside the non-classical regime where such mathematical results are devoid of physical meaning. Be it as it may, the analytical as well as numerical methods that we have introduced here to investigate the transfer functions in Eqs.(20) and (29) are quite general and can be used to analyze many types of variations on the aforementioned equations of motion.

---

[‡‡] Again, for $R \lesssim r_c$, proper accounting for mass-renormalization could change this conclusion.

We close by expanding on our earlier remarks concerning Dirac's relativistic calculation of the self-force acting on an accelerated point-charge.[11] It was pointed out in Sec.1, that Dirac "found a clever way to eliminate the troublesome infinities that had previously hampered the investigations of point particles." Having carefully examined Dirac's argument, we find his "clever way" to *not* be a trick, nor an ad hoc attempt at extracting a meaningful answer from an ill-defined problem in the classical electrodynamics of an accelerated point-charge. As explained in some detail in a two-part paper by one of the authors,[30] Dirac's courageous and unconventional approach — using one-half each of the retarded and advanced potentials — is *inevitable* when one attempts to solve Maxwell's equations for a true point-particle. What is more, Dirac's result can be obtained directly based on the conservation laws of energy and linear momentum, *without* invoking his unconventional assumption.[30] We contend, therefore, that Dirac's relativistic solution to the Abraham-Lorentz problem is indeed the exact solution of Maxwell's equations for an accelerated point-charge.

Dirac's solution for a zero-size particle removes the infinite contribution of the self-force to the particle's mass, making his solution fundamentally different from any exact or approximate solution that may be found for a particle of finite radius $R$. This is why we have refrained from using the notation $R \to 0$ in this paper, lest it imply that our conclusions remain valid when the particle size shrinks to zero. The various appearances throughout the paper of the qualifiers "small," "extremely small," and "exceedingly small" for the particle size are intended to forestall a potential misunderstanding that our results should remain applicable when the particle under consideration reduces to a true (i.e., zero-dimensional) point-charge.

**Acknowledgement**. The authors express their gratitude to Vladimir Hnizdo for commenting on an early draft of this paper, and for generously sharing with them his extensive knowledge of the electrodynamics of charged particles. This work has been supported in part by the AFOSR grant FA9550-19-1-0032.

## Appendix A

The four-dimensional Fourier transform of $\boldsymbol{P}(\boldsymbol{r},t)$ given by Eq.(1) is calculated as follows:

$$\boldsymbol{P}(\boldsymbol{k},\omega) = \int_{-\infty}^{\infty} \boldsymbol{P}(\boldsymbol{r},t) \exp[-i(\boldsymbol{k}\cdot\boldsymbol{r} - \omega t)] \, d\boldsymbol{r} dt$$

$$= P_0\hat{\boldsymbol{z}}\left[\int_{-\infty}^{\infty} \text{sphere}(r/R)e^{-i\boldsymbol{k}\cdot\boldsymbol{r}}d\boldsymbol{r}\right] \times \left[\tfrac{1}{2}\int_{-\infty}^{\infty}(e^{i\omega_s t} + e^{-i\omega_s t})e^{i\omega t}dt\right]$$

$$= \pi P_0\hat{\boldsymbol{z}}[\delta(\omega+\omega_s) + \delta(\omega-\omega_s)] \int_{r=0}^{R}\int_{\theta=0}^{\pi} 2\pi r^2 \sin\theta \exp(-ikr\cos\theta) \, drd\theta$$

$$= 2\pi^2 P_0\hat{\boldsymbol{z}}[\delta(\omega+\omega_s) + \delta(\omega-\omega_s)](-i/k)\int_0^R r\exp(-ikr\cos\theta)|_{\theta=0}^{\pi}dr$$

$$= 2\pi^2 P_0\hat{\boldsymbol{z}}[\delta(\omega+\omega_s) + \delta(\omega-\omega_s)](-i/k)\int_0^R i2r\sin(kr)\,dr$$

$$= 4\pi^2(P_0/k^3)\hat{\boldsymbol{z}}[\delta(\omega+\omega_s) + \delta(\omega-\omega_s)]\int_0^{kR} x\sin x\,dx$$

$$= 4\pi^2(P_0/k^3)\hat{\boldsymbol{z}}[\delta(\omega+\omega_s) + \delta(\omega-\omega_s)]\left(-x\cos x|_{x=0}^{kR} + \int_0^{kR}\cos x\,dx\right)$$

$$= 4\pi^2 R^3 P_0\hat{\boldsymbol{z}}[\delta(\omega+\omega_s) + \delta(\omega-\omega_s)]\,[\sin(kR) - kR\cos(kR)]/(kR)^3$$

$$= 3\pi p_0\hat{\boldsymbol{z}}[\delta(\omega+\omega_s) + \delta(\omega-\omega_s)]\,[\sin(kR) - kR\cos(kR)]/(kR)^3. \tag{A1}$$

In the limit when $kR \to 0$, we find that $[\sin(kR) - kR\cos(kR)]/(kR)^3 \to \tfrac{1}{3}$. The function $\boldsymbol{P}(\boldsymbol{k},\omega)$ is thus well-behaved for all values of $R$ throughout the $(\boldsymbol{k},\omega)$ space.

## Appendix B

The scalar potential $\psi(\boldsymbol{r},t)$ and the vector potential $\boldsymbol{A}(\boldsymbol{r},t)$ are directly computed from the charge- and current-densities $\rho(\boldsymbol{r},t)$ and $\boldsymbol{J}(\boldsymbol{r},t)$, given by Eqs.(3) and (4). In the step-by-step derivations that follow, we seek analytic expressions for the potentials in the regions both inside ($r < R$) and outside ($r > R$) the spherical dipole.

**B1. The vector potential.** In the $(\boldsymbol{r},t)$ spacetime domain The vector potential is obtained by an inverse Fourier integral over the total electric current density distribution, as follows:

$$\boldsymbol{A}(\boldsymbol{r},t) = \frac{\mu_0}{(2\pi)^4}\int_{-\infty}^{\infty}\frac{\boldsymbol{J}(\boldsymbol{k},\omega)}{k^2 - (\omega/c)^2}\exp[i(\boldsymbol{k}\cdot\boldsymbol{r} - \omega t)]\,d\boldsymbol{k}d\omega$$

$$= -\frac{i3\pi\mu_0 p_0\hat{\boldsymbol{z}}}{(2\pi)^4}\int_{-\infty}^{\infty}\frac{\sin(kR) - kR\cos(kR)}{(kR)^3}e^{i\boldsymbol{k}\cdot\boldsymbol{r}}\int_{-\infty}^{\infty}\frac{\omega[\delta(\omega+\omega_s)+\delta(\omega-\omega_s)]}{k^2-(\omega/c)^2}e^{-i\omega t}\,d\omega d\boldsymbol{k}$$

$$= -\frac{3\mu_0 p_0 \omega_s \sin(\omega_s t)\hat{\boldsymbol{z}}}{8\pi^3 R^3}\int_{-\infty}^{\infty}\frac{\sin(kR) - kR\cos(kR)}{k^3[k^2-(\omega_s/c)^2]}e^{i\boldsymbol{k}\cdot\boldsymbol{r}}d\boldsymbol{k}$$

$$= -\frac{3\mu_0 p_0 \omega_s \sin(\omega_s t)\hat{\boldsymbol{z}}}{8\pi^3 R^3}\int_{k=0}^{\infty}\int_{\varphi=0}^{\pi}\frac{\sin(kR) - kR\cos(kR)}{k^3[k^2-(\omega_s/c)^2]}e^{ikr\cos\varphi}2\pi k^2\sin\varphi\,dkd\varphi$$

$$= -\frac{3\mu_0 p_0 \omega_s \sin(\omega_s t)\hat{\boldsymbol{z}}}{4\pi^2 R^3}\int_{k=0}^{\infty}\frac{\sin(kR) - kR\cos(kR)}{k^2[k^2-(\omega_s/c)^2]}\int_{\varphi=0}^{\pi}k\sin\varphi\,e^{ikr\cos\varphi}d\varphi dk$$

$$
\begin{aligned}
&= -\frac{\mathrm{i}3\mu_0 p_0 \omega_s \sin(\omega_s t)\hat{z}}{4\pi^2 R^3 r} \int_{k=0}^{\infty} \frac{\sin(kR)-kR\cos(kR)}{k^2[k^2-(\omega_s/c)^2]} e^{\mathrm{i}kr\cos\varphi}\Big|_{\varphi=0}^{\pi} \mathrm{d}k \\
&= -\frac{3\mu_0 p_0 \omega_s \sin(\omega_s t)\hat{z}}{4\pi^2 R^3 r} \int_{-\infty}^{\infty} \frac{[\sin(kR)-kR\cos(kR)]\sin(kr)}{k^2[k^2-(\omega_s/c)^2]} \mathrm{d}k \\
&= -\frac{\mathrm{i}3\mu_0 p_0 \omega_s \sin(\omega_s t)\hat{z}}{16\pi^2 R^3 r} \int_{-\infty}^{\infty} \frac{[(kR+\mathrm{i})e^{\mathrm{i}kR}+(kR-\mathrm{i})e^{-\mathrm{i}kR}](e^{\mathrm{i}kr}-e^{-\mathrm{i}kr})}{k^2[k^2-(\omega_s/c)^2]} \mathrm{d}k \\
&= -\frac{\mathrm{i}3\mu_0 p_0 \omega_s \sin(\omega_s t)\hat{z}}{16\pi^2 R^3 r} \int_{-\infty}^{\infty} \frac{(kR+\mathrm{i})e^{\mathrm{i}(r+R)k}-(kR-\mathrm{i})e^{-\mathrm{i}(r+R)k}+(kR-\mathrm{i})e^{\mathrm{i}(r-R)k}-(kR+\mathrm{i})e^{-\mathrm{i}(r-R)k}}{k^2[k^2-(\omega_s/c)^2]} \mathrm{d}k \\
&= \frac{3\mu_0 p_0 \omega_s \sin(\omega_s t)\hat{z}}{8\pi^2 R^3 r} \mathrm{Im}\int_{-\infty}^{\infty} \frac{(kR+\mathrm{i})e^{\mathrm{i}(r+R)k}+(kR-\mathrm{i})e^{\mathrm{i}(r-R)k}}{k^2[k-(\omega_s/c)][k+(\omega_s/c)]} \mathrm{d}k.
\end{aligned} \quad (\mathrm{B1})
$$

The preceding integral can be evaluated in the complex plane using Cauchy's theorem. Outside the spherical particle, where $r > R$, both exponential functions appearing in the integrand approach zero on a large semi-circular contour in the upper half-plane. The half-residues at the first-order poles $k = \pm\omega_s/c$ are easy to evaluate. As for the second-order pole at $k = 0$, the product of $k^2$ and the integrand turns out to have a Taylor series expansion around $k = 0$ that has no constant term and no first-order term. As such, the pole at $k = 0$ makes no contribution to the integral. We will have

$$
\begin{aligned}
\boldsymbol{A}_{\mathrm{out}}(\boldsymbol{r},t) &= \frac{3\mu_0 p_0 \omega_s \sin(\omega_s t)\hat{z}}{8\pi^2 R^3 r} \mathrm{Im}\left\{ \mathrm{i}\pi \frac{[(R\omega_s/c)+\mathrm{i}]e^{\mathrm{i}(r+R)\omega_s/c}+[(R\omega_s/c)-\mathrm{i}]e^{\mathrm{i}(r-R)\omega_s/c}}{2(\omega_s/c)^3} \right. \\
&\quad + \mathrm{i}\pi \frac{[(R\omega_s/c)-\mathrm{i}]e^{-\mathrm{i}(r+R)\omega_s/c}+[(R\omega_s/c)+\mathrm{i}]e^{-\mathrm{i}(r-R)\omega_s/c}}{2(\omega_s/c)^3} + \mathrm{i}\pi \frac{\mathrm{d}}{\mathrm{d}k}\frac{(kR+\mathrm{i})e^{\mathrm{i}(r+R)k}+(kR-\mathrm{i})e^{\mathrm{i}(r-R)k}}{k^2-(\omega_s/c)^2}\Big|_{k=0}^{\;\;0} \bigg\} \\
&= \frac{3\mu_0 p_0 \omega_s \sin(\omega_s t)\hat{z}}{8\pi R^3 r} \left\{ \frac{(R\omega_s/c)\cos[(r+R)\omega_s/c]-\sin[(r+R)\omega_s/c]}{(\omega_s/c)^3} \right. \\
&\quad + \left. \frac{(R\omega_s/c)\cos[(r-R)\omega_s/c]+\sin[(r-R)\omega_s/c]}{(\omega_s/c)^3} \right\} \\
&= -\left(\frac{3\mu_0 p_0 \omega_s \hat{z}}{4\pi r}\right) \frac{[\sin(R\omega_s/c)-(R\omega_s/c)\cos(R\omega_s/c)]\cos(r\omega_s/c)\sin(\omega_s t)}{(R\omega_s/c)^3}.
\end{aligned} \quad (\mathrm{B2})
$$

A similar method applies to Eq.(B1) inside the spherical particle ($r < R$), except that the second term of the integrand now vanishes on a large semi-circular contour in the *lower* half of the complex plane. Also, this time the 2$^{\mathrm{nd}}$ order pole at $k = 0$ does make a contribution. We have

$$
\begin{aligned}
\boldsymbol{A}_{\mathrm{in}}(\boldsymbol{r},t) &= \frac{3\mu_0 p_0 \omega_s \sin(\omega_s t)\hat{z}}{8\pi^2 R^3 r} \mathrm{Im}\left\{ \mathrm{i}\pi \frac{[(R\omega_s/c)+\mathrm{i}]e^{\mathrm{i}(r+R)\omega_s/c}-[(R\omega_s/c)-\mathrm{i}]e^{\mathrm{i}(r-R)\omega_s/c}}{2(\omega_s/c)^3} \right. \\
&\quad + \mathrm{i}\pi \frac{[(R\omega_s/c)-\mathrm{i}]e^{-\mathrm{i}(r+R)\omega_s/c}-[(R\omega_s/c)+\mathrm{i}]e^{-\mathrm{i}(r-R)\omega_s/c}}{2(\omega_s/c)^3} \\
&\quad + \mathrm{i}\pi \frac{\mathrm{d}}{\mathrm{d}k}\frac{(kR+\mathrm{i})e^{\mathrm{i}(r+R)k}-(kR-\mathrm{i})e^{\mathrm{i}(r-R)k}}{k^2-(\omega_s/c)^2}\bigg|_{k=0} \bigg\} \\
&= \frac{3\mu_0 p_0 \omega_s \sin(\omega_s t)\hat{z}}{8\pi R^3 r}\left\{\frac{(R\omega_s/c)\cos[(r+R)\omega_s/c]-\sin[(r+R)\omega_s/c]}{(\omega_s/c)^3}\right.
\end{aligned}
$$

Note that the first term in the Taylor series expansion around $k = 0$ is zero.

$$\left.\underbrace{\begin{array}{l}\sin(a\pm b)=\sin a\cos b\pm\cos a\sin b\\ \cos(a\pm b)=\cos a\cos b\mp\sin a\sin b\end{array}}-\frac{(R\omega_s/c)\cos[(r-R)\omega_s/c]+\sin[(r-R)\omega_s/c]}{(\omega_s/c)^3}+\frac{2r}{(\omega_s/c)^2}\right\}$$

$$=\frac{3\mu_0 p_0\omega_s\hat{\mathbf{z}}}{4\pi(R\omega_s/c)^3 r}\{(r\omega_s/c)-[\cos(R\omega_s/c)+(R\omega_s/c)\sin(R\omega_s/c)]\sin(r\omega_s/c)\}\sin(\omega_s t). \quad \text{(B3)}$$

Now, in the far field, the vector potential $\mathbf{A}_{\text{out}}(\mathbf{r},t)$ of Eq.(B2) contains both retarded terms in the form of $\sin[\omega_s(t-r/c)]$ and advanced terms in the form of $\sin[\omega_s(t+r/c)]$. To eliminate the advanced terms, the heretofore neglected contributions of source-free (i.e., vacuum) terms must be added to both $\mathbf{A}_{\text{out}}$ of Eq.(B2) and $\mathbf{A}_{\text{in}}$ of Eq.(B3). We will return to this task later, after computing the scalar potential in the next section.

**B2. The scalar potential**. The scalar potential in the spacetime domain is obtained similarly, via an inverse Fourier transformation, as follows:

$$\psi(\mathbf{r},t)=(2\pi)^{-4}\int_{-\infty}^{\infty}\frac{\rho(\mathbf{k},\omega)}{\varepsilon_0[k^2-(\omega/c)^2]}\exp[\mathrm{i}(\mathbf{k}\cdot\mathbf{r}-\omega t)]\mathrm{d}\mathbf{k}\mathrm{d}\omega$$

$$=-\frac{\mathrm{i}3\pi p_0\hat{\mathbf{z}}}{(2\pi)^4\varepsilon_0}\cdot\int_{-\infty}^{\infty}\frac{\mathbf{k}[\sin(kR)-kR\cos(kR)]\exp(\mathrm{i}\mathbf{k}\cdot\mathbf{r})}{(kR)^3}\int_{-\infty}^{\infty}\frac{[\delta(\omega+\omega_s)+\delta(\omega-\omega_s)]\exp(-\mathrm{i}\omega t)}{k^2-(\omega/c)^2}\mathrm{d}\omega\,\mathrm{d}\mathbf{k}$$

$$=\frac{\mathrm{i}3p_0\cos(\omega_s t)\hat{\mathbf{z}}}{8\pi^3\varepsilon_0}\cdot\int_{-\infty}^{\infty}\frac{\mathbf{k}[kR\cos(kR)-\sin(kR)]\exp(\mathrm{i}\mathbf{k}\cdot\mathbf{r})}{(kR)^3[k^2-(\omega_s/c)^2]}\mathrm{d}\mathbf{k}$$

$$=\frac{\mathrm{i}3p_0\cos(\omega_s t)\hat{\mathbf{z}}}{8\pi^3\varepsilon_0}\cdot\int_{k=0}^{\infty}\int_{\varphi=0}^{\pi}\frac{(k\hat{\mathbf{r}}\cos\varphi)[kR\cos(kR)-\sin(kR)]\exp(\mathrm{i}kr\cos\varphi)}{(kR)^3[k^2-(\omega_s/c)^2]}2\pi k^2\sin\varphi\,\mathrm{d}k\mathrm{d}\varphi$$

$$=\frac{\mathrm{i}3p_0\cos(\omega_s t)(\hat{\mathbf{z}}\cdot\hat{\mathbf{r}})}{4\pi^2\varepsilon_0}\int_{k=0}^{\infty}\frac{kR\cos(kR)-\sin(kR)}{R^3[k^2-(\omega_s/c)^2]}\int_{\varphi=0}^{\pi}\sin\varphi\cos\varphi\exp(\mathrm{i}kr\cos\varphi)\,\mathrm{d}\varphi\,\mathrm{d}k$$

$$=\frac{3p_0\cos(\omega_s t)\cos\theta}{2\pi^2\varepsilon_0}\int_0^{\infty}\frac{kR\cos(kR)-\sin(kR)}{R^3[k^2-(\omega_s/c)^2]}\times\frac{kr\cos(kr)-\sin(kr)}{(kr)^2}\mathrm{d}k$$

$$=\frac{3p_0\cos(\omega_s t)\cos\theta}{8\pi^2\varepsilon_0 R^3 r^2}\int_0^{\infty}\frac{[(kR+\mathrm{i})e^{\mathrm{i}kR}+(kR-\mathrm{i})e^{-\mathrm{i}kR}]\times[(kr+\mathrm{i})e^{\mathrm{i}kr}+(kr-\mathrm{i})e^{-\mathrm{i}kr}]}{k^2[k^2-(\omega_s/c)^2]}\mathrm{d}k$$

$$=\frac{3p_0\cos(\omega_s t)\cos\theta}{16\pi^2\varepsilon_0 R^3 r^2}\left\{\int_{-\infty}^{\infty}\frac{[Rrk^2-1+\mathrm{i}(r+R)k]e^{\mathrm{i}(r+R)k}+[Rrk^2-1-\mathrm{i}(r+R)k]e^{-\mathrm{i}(r+R)k}}{k^2[k^2-(\omega_s/c)^2]}\mathrm{d}k\right.$$

$$\left.+\int_{-\infty}^{\infty}\frac{[Rrk^2+1-\mathrm{i}(r-R)k]e^{\mathrm{i}(r-R)k}+[Rrk^2+1+\mathrm{i}(r-R)k]e^{-\mathrm{i}(r-R)k}}{k^2[k^2-(\omega_s/c)^2]}\mathrm{d}k\right\}$$

$$=\frac{3p_0\cos(\omega_s t)\cos\theta}{8\pi^2\varepsilon_0 R^3 r^2}\operatorname{Re}\int_{-\infty}^{\infty}\frac{[Rrk^2-1+\mathrm{i}(r+R)k]e^{\mathrm{i}(r+R)k}+[Rrk^2+1-\mathrm{i}(r-R)k]e^{\mathrm{i}(r-R)k}}{k^2[k-(\omega_s/c)][k+(\omega_s/c)]}\mathrm{d}k. \quad \text{(B4)}$$

Once again, the preceding integral can be evaluated in the complex plane using Cauchy's theorem. Outside the spherical particle, where $r>R$, both exponential functions appearing in the integrand approach zero on a large semi-circular contour in the upper half-plane. The half-residues at the first-order poles $k=\pm\omega_s/c$ are easy to evaluate. As for the second-order pole at $k=0$, the product of $k^2$ and the integrand turns out to have a Taylor series expansion around $k = 0$ that has no constant term and no first-order term. As such, the pole at $k = 0$ makes no contribution to the integral. We will have

$$\psi_{\text{out}}(\boldsymbol{r}, t) = \frac{3p_0 \cos(\omega_s t) \cos\theta}{8\pi^2 \varepsilon_0 R^3 r^2}$$

$$\times \text{Re}\left\{ i\pi \frac{[Rr(\omega_s/c)^2 - 1 + i(r+R)(\omega_s/c)]e^{i(r+R)\omega_s/c} + [Rr(\omega_s/c)^2 + 1 - i(r-R)(\omega_s/c)]e^{i(r-R)\omega_s/c}}{2(\omega_s/c)^3} \right.$$

$$-i\pi \frac{[Rr(\omega_s/c)^2 - 1 - i(r+R)(\omega_s/c)]e^{-i(r+R)\omega_s/c} + [Rr(\omega_s/c)^2 + 1 + i(r-R)(\omega_s/c)]e^{-i(r-R)\omega_s/c}}{2(\omega_s/c)^3}$$

$$+ i\pi \frac{d}{dk} \left. \frac{[Rrk^2 - 1 + i(r+R)k]e^{i(r+R)k} + [Rrk^2 + 1 - i(r-R)k]e^{i(r-R)k}}{k^2 - (\omega_s/c)^2} \right|_{k=0} \xrightarrow{0} \Bigg\}$$

$$= -\frac{3p_0 \cos(\omega_s t) \cos\theta}{8\pi\varepsilon_0 R^3 r^2} \left\{ \frac{[Rr(\omega_s/c)^2 - 1]\sin[(r+R)\omega_s/c] + (r+R)(\omega_s/c)\cos[(r+R)\omega_s/c]}{(\omega_s/c)^3} \right.$$

$\sin(a \pm b) = \sin a \cos b \pm \cos a \sin b$
$\cos(a \pm b) = \cos a \cos b \mp \sin a \sin b$

$$+ \frac{[Rr(\omega_s/c)^2 + 1]\sin[(r-R)\omega_s/c] - (r-R)(\omega_s/c)\cos[(r-R)\omega_s/c]}{(\omega_s/c)^3} \Bigg\}$$

$$= \left(\frac{3p_0 \cos\theta}{4\pi\varepsilon_0 r^2}\right) \frac{[\sin(R\omega_s/c) - (R\omega_s/c)\cos(R\omega_s/c)] \times [\cos(r\omega_s/c) + (r\omega_s/c)\sin(r\omega_s/c)]\cos(\omega_s t)}{(R\omega_s/c)^3}. \quad (B5)$$

A similar method applies to Eq.(B4) inside the spherical particle ($r < R$), except that the second term of the integrand now vanishes on a large semi-circular contour in the *lower* half of the complex plane. Again, the 2$^{\text{nd}}$ order pole at $k = 0$ fails to make a contribution. We will have

$$\psi_{\text{in}}(\boldsymbol{r}, t) = \frac{3p_0 \cos(\omega_s t) \cos\theta}{8\pi^2 \varepsilon_0 R^3 r^2}$$

$$\times \text{Re}\left\{ i\pi \frac{[Rr(\omega_s/c)^2 - 1 + i(r+R)(\omega_s/c)]e^{i(r+R)\omega_s/c} - [Rr(\omega_s/c)^2 + 1 - i(r-R)(\omega_s/c)]e^{i(r-R)\omega_s/c}}{2(\omega_s/c)^3} \right.$$

$$-i\pi \frac{[Rr(\omega_s/c)^2 - 1 - i(r+R)(\omega_s/c)]e^{-i(r+R)\omega_s/c} - [Rr(\omega_s/c)^2 + 1 + i(r-R)(\omega_s/c)]e^{-i(r-R)\omega_s/c}}{2(\omega_s/c)^3}$$

$$+ i\pi \frac{d}{dk} \left. \frac{[Rrk^2 - 1 + i(r+R)k]e^{i(r+R)k} - [Rrk^2 + 1 - i(r-R)k]e^{i(r-R)k}}{k^2 - (\omega_s/c)^2} \right|_{k=0} \xrightarrow{0} \Bigg\}$$

$$= -\frac{3p_0 \cos(\omega_s t) \cos\theta}{8\pi\varepsilon_0 R^3 r^2} \left\{ \frac{[Rr(\omega_s/c)^2 - 1]\sin[(r+R)\omega_s/c] + (r+R)(\omega_s/c)\cos[(r+R)\omega_s/c]}{(\omega_s/c)^3} \right.$$

$\sin(a \pm b) = \sin a \cos b \pm \cos a \sin b$
$\cos(a \pm b) = \cos a \cos b \mp \sin a \sin b$

$$- \frac{[Rr(\omega_s/c)^2 + 1]\sin[(r-R)\omega_s/c] - (r-R)(\omega_s/c)\cos[(r-R)\omega_s/c]}{(\omega_s/c)^3} \Bigg\}$$

$$= \left(\frac{3p_0 r \cos\theta}{4\pi\varepsilon_0 R^3}\right) \frac{[\cos(R\omega_s/c) + (R\omega_s/c)\sin(R\omega_s/c)] \times [\sin(r\omega_s/c) - (r\omega_s/c)\cos(r\omega_s/c)]\cos(\omega_s t)}{(r\omega_s/c)^3}. \quad (B6)$$

As before, the scalar potential $\psi_{\text{out}}(\boldsymbol{r}, t)$ of Eq.(B5) contains both advanced and retarded terms in the form of $\cos[\omega_s(t \pm r/c)]$ and $\sin[\omega_s(t \pm r/c)]$. To eliminate the advanced terms, the heretofore neglected contributions of source-free (i.e., vacuum) terms must be added to both $\psi_{\text{out}}$ of Eq.(B5) and $\psi_{\text{in}}$ of Eq.(B6). The vacuum potentials are computed in the following section, after which we will return to derive the complete expressions for the scalar and vector potentials of our spherical dipole.

**B3. Computing the vacuum potentials.** The vacuum vector potential $\boldsymbol{A}_{\text{vac}}(\boldsymbol{r},t)$ is a superposition of plane-waves propagating in free space, having frequencies $\pm\omega_s$, $k$-vectors $\boldsymbol{k} = (\omega_s/c)\hat{\boldsymbol{k}}$, and the overall Fourier domain distribution

$$\boldsymbol{A}_{\text{vac}}(\boldsymbol{k},\omega) = A_0\hat{\boldsymbol{z}}\delta[k - (\omega_s/c)][\delta(\omega - \omega_s) + \delta(\omega + \omega_s)]. \tag{B7}$$

The constant coefficient $A_0$ will be determined further below. The spacetime profile of the above distribution is found straightforwardly via the following inverse Fourier transformation:

$$\boldsymbol{A}_{\text{vac}}(\boldsymbol{r},t) = (2\pi)^{-4}\int_{-\infty}^{\infty} A_0\hat{\boldsymbol{z}}\delta[k - (\omega_s/c)][\delta(\omega - \omega_s) + \delta(\omega + \omega_s)]\exp[i(\boldsymbol{k}\cdot\boldsymbol{r} - \omega t)]\,\mathrm{d}\boldsymbol{k}\mathrm{d}\omega$$

$$= (2\pi)^{-4}A_0\hat{\boldsymbol{z}}\int_{-\infty}^{\infty}\delta[k - (\omega_s/c)]e^{i\boldsymbol{k}\cdot\boldsymbol{r}}\mathrm{d}\boldsymbol{k}\int_{-\infty}^{\infty}[\delta(\omega - \omega_s) + \delta(\omega + \omega_s)]e^{-i\omega t}\mathrm{d}\omega$$

$$= 2(2\pi)^{-4}A_0\hat{\boldsymbol{z}}\cos(\omega_s t)\int_{-\infty}^{\infty}\delta[k - (\omega_s/c)]e^{i\boldsymbol{k}\cdot\boldsymbol{r}}\mathrm{d}\boldsymbol{k}$$

$$= 2(2\pi)^{-4}A_0\hat{\boldsymbol{z}}\cos(\omega_s t)\int_{k=0}^{\infty}\int_{\varphi=0}^{\pi}\delta[k - (\omega_s/c)]\,e^{ikr\cos\varphi}2\pi k^2\sin\varphi\,\mathrm{d}k\mathrm{d}\varphi$$

$$= 2(2\pi)^{-3}A_0\hat{\boldsymbol{z}}\cos(\omega_s t)\int_{k=0}^{\infty}\delta[k - (\omega_s/c)]k^2\int_{\varphi=0}^{\pi}\sin\varphi\,e^{ikr\cos\varphi}\mathrm{d}\varphi\mathrm{d}k$$

$$= 2(2\pi)^{-3}A_0\hat{\boldsymbol{z}}\cos(\omega_s t)\int_0^{\infty}\delta[k - (\omega_s/c)]k^2[2\sin(kr)/(kr)]\mathrm{d}k$$

$$= \left(\frac{\omega_s A_0\hat{\boldsymbol{z}}}{2\pi^3 cr}\right)\sin(r\omega_s/c)\cos(\omega_s t). \tag{B8}$$

Comparison with Eq.(B2) reveals that, in order to eliminate the advanced term in the overall vector potential, the coefficient $A_0$ of the vacuum potential must be set to

$$A_0 = \left(\frac{3\pi^2 c\mu_0 p_0}{2}\right)\frac{\sin(R\omega_s/c) - (R\omega_s/c)\cos(R\omega_s/c)}{(R\omega_s/c)^3}. \tag{B9}$$

We thus arrive at the desired form of the vacuum vector potential, namely,

$$\boldsymbol{A}_{\text{vac}}(\boldsymbol{r},t) = \left(\frac{3\mu_0\omega_s p_0\hat{\boldsymbol{z}}}{4\pi r}\right)\left[\frac{\sin(R\omega_s/c) - (R\omega_s/c)\cos(R\omega_s/c)}{(R\omega_s/c)^3}\right]\sin(r\omega_s/c)\cos(\omega_s t). \tag{B10}$$

The corresponding vacuum scalar potential $\psi_{\text{vac}}(\boldsymbol{r},t)$ is a superposition of plane-waves in free space, having frequencies $\pm\omega_s$, $k$-vectors $\boldsymbol{k} = (\omega_s/c)\hat{\boldsymbol{k}}$, and the Fourier domain distribution

$$\psi_{\text{vac}}(\boldsymbol{k},\omega) = \psi_0\delta[k - (\omega_s/c)][\delta(\omega - \omega_s) - \delta(\omega + \omega_s)]\hat{\boldsymbol{k}}\cdot\hat{\boldsymbol{z}}. \tag{B11}$$

The constant coefficient $\psi_0$ will be determined further below. The spacetime profile of the above distribution is found straightforwardly via inverse Fourier transformation, as follows:

$$\psi_{\text{vac}}(\boldsymbol{r},t) = (2\pi)^{-4}\int_{-\infty}^{\infty}\psi_0\delta[k - (\omega_s/c)][\delta(\omega - \omega_s) - \delta(\omega + \omega_s)]\hat{\boldsymbol{k}}\cdot\hat{\boldsymbol{z}}\exp[i(\boldsymbol{k}\cdot\boldsymbol{r} - \omega t)]\,\mathrm{d}\boldsymbol{k}\mathrm{d}\omega$$

$$= (2\pi)^{-4}\psi_0\hat{\boldsymbol{z}}\cdot\int_{-\infty}^{\infty}\delta[k - (\omega_s/c)]\hat{\boldsymbol{k}}e^{i\boldsymbol{k}\cdot\boldsymbol{r}}\mathrm{d}\boldsymbol{k}\int_{-\infty}^{\infty}[\delta(\omega - \omega_s) - \delta(\omega + \omega_s)]e^{-i\omega t}\mathrm{d}\omega$$

$$= -2\mathrm{i}(2\pi)^{-4}\psi_0\sin(\omega_s t)\,\hat{\boldsymbol{z}}\cdot\int_{-\infty}^{\infty}\delta[k - (\omega_s/c)]\hat{\boldsymbol{k}}e^{i\boldsymbol{k}\cdot\boldsymbol{r}}\mathrm{d}\boldsymbol{k}$$

$$= -2\mathrm{i}(2\pi)^{-4}\psi_0\sin(\omega_s t)\,\hat{\boldsymbol{z}}\cdot\int_{k=0}^{\infty}\int_{\varphi=0}^{\pi}\delta[k - (\omega_s/c)]\cos\varphi\,\hat{\boldsymbol{r}}\,e^{ikr\cos\varphi}2\pi k^2\sin\varphi\,\mathrm{d}k\mathrm{d}\varphi$$

$$= -2\mathrm{i}(2\pi)^{-3}\psi_0 \sin(\omega_s t)\,\hat{\mathbf{z}}\cdot\hat{\mathbf{r}} \int_{k=0}^{\infty} \delta[k-(\omega_s/c)]k^2 \int_{\varphi=0}^{\pi} \sin\varphi\cos\varphi\, e^{\mathrm{i}kr\cos\varphi}\,\mathrm{d}\varphi\,\mathrm{d}k$$

$$= 4(2\pi)^{-3}\psi_0 \sin(\omega_s t)\cos\theta \int_0^{\infty} \delta[k-(\omega_s/c)]k^2 \left[\frac{\sin(kr) - kr\cos(kr)}{(kr)^2}\right]\mathrm{d}k$$

$$= \frac{\psi_0 \cos\theta}{2\pi^3 r^2}[\sin(r\omega_s/c) - (r\omega_s/c)\cos(r\omega_s/c)]\sin(\omega_s t). \tag{B12}$$

Comparison with Eq.(B5) reveals that, in order to eliminate the advanced terms in the overall scalar potential, the coefficient $\psi_0$ must be set to

$$\psi_0 = \left(\frac{3\pi^2 p_0}{2\varepsilon_0}\right)\frac{\sin(R\omega_s/c) - (R\omega_s/c)\cos(R\omega_s/c)}{(R\omega_s/c)^3}. \tag{B13}$$

We thus arrive at the desired vacuum scalar potential, namely,

$$\psi_{\mathrm{vac}}(\mathbf{r},t) = \left(\frac{3p_0 \cos\theta}{4\pi\varepsilon_0 r^2}\right)\frac{[\sin(R\omega_s/c) - (R\omega_s/c)\cos(R\omega_s/c)][\sin(r\omega_s/c) - (r\omega_s/c)\cos(r\omega_s/c)]\sin(\omega_s t)}{(R\omega_s/c)^3}. \tag{B14}$$

Upon adding the vacuum potentials of Eqs.(B10) and (B14) to the dipolar potentials found previously in Sections B1 and B2, the terms corresponding to advanced potentials are eliminated, leaving the exterior fields of the dipole with retarded terms only.

---

**Digression**: The Lorenz gauge, $\boldsymbol{\nabla}\cdot\mathbf{A}(\mathbf{r},t) + (1/c^2)\,\partial\psi(\mathbf{r},t)/\partial t = 0$, must be satisfied by the vacuum potentials of Eqs.(B10) and (B14). To confirm this, note that the potentials in the Fourier domain are given by

$$\mathbf{A}_{\mathrm{vac}}(\mathbf{k},\omega) = \left(\frac{3\pi^2 c\mu_0 p_0}{2}\right)\left[\frac{\sin(R\omega_s/c) - (R\omega_s/c)\cos(R\omega_s/c)}{(R\omega_s/c)^3}\right]\delta[k-(\omega_s/c)][\delta(\omega-\omega_s)+\delta(\omega+\omega_s)]\hat{\mathbf{z}}, \tag{B15}$$

$$\psi_{\mathrm{vac}}(\mathbf{k},\omega) = \left(\frac{3\pi^2 p_0}{2\varepsilon_0}\right)\left[\frac{\sin(R\omega_s/c) - (R\omega_s/c)\cos(R\omega_s/c)}{(R\omega_s/c)^3}\right]\delta[k-(\omega_s/c)][\delta(\omega-\omega_s)-\delta(\omega+\omega_s)]\hat{\mathbf{k}}\cdot\hat{\mathbf{z}}. \tag{B16}$$

These expressions clearly satisfy the Fourier version of the Lorenz gauge, namely $\mathbf{k}\cdot\mathbf{A}(\mathbf{k},\omega) = (\omega/c^2)\psi(\mathbf{k},\omega)$.

---

**B4. Complete expressions of the vector and scalar potentials**. Adding the vacuum vector potential of Eq.(B10) to Eqs.(B2) and (B3) now yields the correct (i.e., retarded) form of the vector potential function both inside and outside the spherical particle, as follows:

$$\mathbf{A}_{\mathrm{out}}(\mathbf{r},t) = -\left(\frac{3\mu_0 \omega_s p_0 \hat{\mathbf{z}}}{4\pi r}\right)\left[\frac{\sin(R\omega_s/c) - (R\omega_s/c)\cos(R\omega_s/c)}{(R\omega_s/c)^3}\right]\sin[\omega_s(t - r/c)]. \tag{B17}$$

$$\mathbf{A}_{\mathrm{in}}(\mathbf{r},t) = \frac{3\mu_0 \omega_s p_0 \hat{\mathbf{z}}}{4\pi(R\omega_s/c)^3 r}\{[\sin(R\omega_s/c) - (R\omega_s/c)\cos(R\omega_s/c)]\sin(r\omega_s/c)\cos(\omega_s t)$$
$$-\{[\cos(R\omega_s/c) + (R\omega_s/c)\sin(R\omega_s/c)]\sin(r\omega_s/c) - (r\omega_s/c)\}\sin(\omega_s t)\}. \tag{B18}$$

Similarly, adding the vacuum scalar potential of Eq.(B14) to Eqs.(B5) and (B6) yields the retarded form of the scalar potential inside as well as outside the particle; that is,

$$\psi_{\mathrm{out}}(\mathbf{r},t) = \left(\frac{3p_0 \cos\theta}{4\pi\varepsilon_0 r^2}\right)\left\{\frac{[\sin(R\omega_s/c) - (R\omega_s/c)\cos(R\omega_s/c)]\times[\cos(r\omega_s/c) + (r\omega_s/c)\sin(r\omega_s/c)]\cos(\omega_s t)}{(R\omega_s/c)^3}\right.$$
$$\left. + \frac{[\sin(R\omega_s/c) - (R\omega_s/c)\cos(R\omega_s/c)]\times[\sin(r\omega_s/c) - (r\omega_s/c)\cos(r\omega_s/c)]\sin(\omega_s t)}{(R\omega_s/c)^3}\right\}$$

$$= \left(\frac{3p_0 \cos\theta}{4\pi\varepsilon_0 r^2}\right)\frac{[\sin(R\omega_s/c) - (R\omega_s/c)\cos(R\omega_s/c)]\times\{\cos[\omega_s(t-r/c)] - (r\omega_s/c)\sin[\omega_s(t-r/c)]\}}{(R\omega_s/c)^3}. \tag{B19}$$

$$\psi_{\text{in}}(\boldsymbol{r},t) = \left(\frac{3p_0 r\cos\theta}{4\pi\varepsilon_0 R^3}\right)\bigg\{\frac{[\cos(R\omega_s/c)+(R\omega_s/c)\sin(R\omega_s/c)]\times[\sin(r\omega_s/c)-(r\omega_s/c)\cos(r\omega_s/c)]\cos(\omega_s t)}{(r\omega_s/c)^3}$$
$$+\frac{[\sin(R\omega_s/c)-(R\omega_s/c)\cos(R\omega_s/c)]\times[\sin(r\omega_s/c)-(r\omega_s/c)\cos(r\omega_s/c)]\sin(\omega_s t)}{(r\omega_s/c)^3}\bigg\}$$
$$= \left(\frac{3p_0 r\cos\theta}{4\pi\varepsilon_0 R^3}\right)\times\left[\frac{\sin(r\omega_s/c)-(r\omega_s/c)\cos(r\omega_s/c)}{(r\omega_s/c)^3}\right]$$
$$\times\{[\cos(R\omega_s/c)+(R\omega_s/c)\sin(R\omega_s/c)]\cos(\omega_s t)$$
$$+[\sin(R\omega_s/c)-(R\omega_s/c)\cos(R\omega_s/c)]\sin(\omega_s t)\}. \tag{B20}$$

These scalar and vector potentials may now be used to evaluate the electric and magnetic fields inside as well as outside the spherical dipole.

## Appendix C

In Sec.5, we tracked the poles of the transfer function of a spherical dipole in the complex $\omega$-plane and argued that the dipole's impulse-response must be causal. Here, we demonstrate the conservation of energy by showing that, in the absence of non-radiative damping (i.e., when $\gamma = 0$), the dipole, which instantaneously acquires its energy at $t = 0$ from the externally applied impulsive $E$-field, releases all of this energy into the surrounding space via EM radiation.

Let the response of the dipole to the impulsive $E$-field $\boldsymbol{E}(t) = \delta(t)\hat{\boldsymbol{z}}$ be denoted by $\wp(t)\hat{\boldsymbol{z}}$, and let $\wp(\omega) = \int_{-\infty}^{\infty} \wp(t)e^{i\omega t}dt$ be the Fourier transform of $\wp(t)$. Suppose $\wp(\omega)$ is sampled at regular intervals of $\Delta\omega$, corresponding to a set of discrete dipoles having amplitudes $p_0 = 2|\wp(\omega)|\Delta\omega/2\pi$, each oscillating at the respective frequency $\omega$, which is an integer-multiple of $\Delta\omega$. The radiated power is thus the sum of Eq.(16) over all such single-frequency oscillators across a range of $\omega$ that extends from 0 to $\infty$. The periodic function $\wp(t)$ obtained by a superposition of these discrete frequency dipoles thus has the period $T = 2\pi/\Delta\omega$. Multiplying the radiation rate of each oscillator with the repetition period $T$ yields the EM energy that the individual dipole oscillator emits during the period $T$. In the limit when $\Delta\omega \to 0$, we will have $Tp_0^2 \to (2/\pi)|\wp(\omega)|^2 d\omega$ and, therefore, the total radiated EM energy $\mathcal{E}$ in response to the impulsive excitation can be computed by integrating Eq.(16) over all frequencies $\omega$; that is,

$$\mathcal{E} = \frac{2}{\pi}\int_0^\infty \left(\frac{3|\wp(\omega)|}{4\pi R^3}\right)^2 \frac{4\pi\omega}{3\varepsilon_0(\omega/c)^3}[\sin(R\omega/c)-(R\omega/c)\cos(R\omega/c)]^2 d\omega. \tag{C1}$$

Considering that the Fourier spectrum of $\boldsymbol{E}(t) = \delta(t)\hat{\boldsymbol{z}}$ is unity at all frequencies $\omega$ ranging from $-\infty$ to $\infty$, the transfer function $p_0/E_0$ of the dipole is given by Eq.(20), as follows:

$$\wp(\omega) = \frac{q^2/m_0}{\omega_0^2 - \omega^2 - \omega_p^2\Gamma(\omega) - i\gamma\omega}. \tag{C2}$$

Here $\omega_p^2 = q^2/(\varepsilon_0 m_0 v)$, with $v = 4\pi R^3/3$ being the volume of the dipole and, from Eq.(19),

$$\Gamma(\omega) = \frac{2[\sin(R\omega/c) - (R\omega/c)\cos(R\omega/c)]\times[1 - i(R\omega/c)]e^{iR\omega/c}}{(R\omega/c)^3} - 1. \tag{C3}$$

The impulse-response $\wp(t)$ is the inverse Fourier transform of the transfer function; that is,

$$\wp(t) = \frac{1}{2\pi}\int_{-\infty}^{\infty}\frac{q^2/m_0}{\omega_0^2 - \omega^2 - \omega_p^2\Gamma(\omega) - i\gamma\omega}e^{-i\omega t}d\omega. \tag{C4}$$

Since all the poles of $\wp(\omega)$ reside in the lower-half of the complex $\omega$-plane, $\wp(t) = 0$ for $t \leq 0$. The vanishing of $\wp(0)$ can be numerically verified by showing that the total area under the function $\wp(\omega)$ is zero. In what follows, we shall set $\gamma = 0$ to eliminate non-radiative losses, thus focusing our attention exclusively on the radiated EM energy. The initial slope $\dot{\wp}(t)|_{t=0}$ of the impulse-response function is obtained by differentiating Eq.(C4) with respect to time, then setting $t = 0$. We find

$$\dot{\wp}(t)|_{t=0} = -\frac{i}{2\pi}\int_{-\infty}^{\infty}\frac{(q^2/m_0)\omega}{\omega_0^2-\omega^2-\omega_p^2\Gamma(\omega)}d\omega = \frac{1}{2\pi}\int_{-\infty}^{\infty}\frac{\omega_p^2\mathrm{Im}[\Gamma(\omega)]\,(q^2/m_0)\omega}{|\omega_0^2-\omega^2-\omega_p^2\,\Gamma(\omega)|^2}d\omega$$

$$= \frac{1}{2\pi}\int_{-\infty}^{\infty}\frac{(q^2/m_0)^2\omega}{\varepsilon_0 v|\omega_0^2-\omega^2-\omega_p^2\Gamma(\omega)|^2} \times \frac{2[\sin(R\omega/c)-(R\omega/c)\cos(R\omega/c)]^2}{(R\omega/c)^3}d\omega$$

$$= \frac{2}{\pi}\int_0^{\infty}\frac{|\wp(\omega)|^2}{v^2} \times \frac{4\pi\omega}{3\varepsilon_0(\omega/c)^3}[\sin(R\omega/c)-(R\omega/c)\cos(R\omega/c)]^2 d\omega. \tag{C5}$$

Considering that the rate of transfer of EM energy from the applied field $\boldsymbol{E}(t)$ to the dipole is $\boldsymbol{E}(t)\cdot\dot{\boldsymbol{\wp}}(t)$, and that the overall energy picked up by the dipole is $\int_{-\infty}^{\infty}\boldsymbol{E}(t)\cdot\dot{\boldsymbol{\wp}}(t)\mathrm{d}t = \dot{\wp}(0)$, a comparison of Eq.(C5) with Eq.(C1) confirms that the energy that is instantaneously picked up by the dipole at $t = 0$, is subsequently radiated away in the form of EM radiation.